\documentclass[
 article,
 groupedaddress,
 showpacs,
 preprintnumbers,
 amsmath,
 amssymb,
 aps,
 prstab,
 floatfix,
]{revtex4-1}

\usepackage{graphicx}					
\usepackage{verbatim}					
\usepackage{dcolumn}                    
\usepackage[ruled,vlined]{algorithm2e}
\include{pythonlisting}
\usepackage{color}

\newtheorem{proposition}{Proposition}[subsection]

\newcommand{\cs}{{\mathrm{C.S.}}}

\newcommand{\sn}{\mathrm{sn}}
\newcommand{\cn}{\mathrm{cn}}
\newcommand{\dn}{\mathrm{dn}}
\newcommand{\asn}{\mathrm{arcsn}}
\newcommand{\acn}{\mathrm{arccn}}
\newcommand{\adn}{\mathrm{arcdn}}
\newcommand{\ads}{\mathrm{arcds}}
\newcommand{\acs}{\mathrm{arccs}}
\newcommand{\ans}{\mathrm{arcns}}

\newcommand{\J}{\mathrm{J}}
\newcommand{\Tr}{\mathrm{Tr}}

\newcommand{\h}{\mathrm{H}}

\newcommand{\const}{\mathrm{const}}

\newcommand{\ds}{\displaystyle}

\newcommand{\dd}{\mathrm{d}}
\newcommand{\pd}{\partial}

\newcommand{\K}{\mathcal{K}}

\begin{document}

\title{ McMillan map 
        and nonlinear Twiss parameters}
\author{T.~Zolkin}
\email{zolkin@fnal.gov}
\affiliation{Fermilab, PO Box 500, Batavia, IL 60510-5011}
\author{S.~Nagaitsev}
\email{nsergei@fnal.gov}
\affiliation{Fermilab, PO Box 500, Batavia, IL 60510-5011}
\affiliation{The University of Chicago, Chicago, IL 60637}
\author{I.~Morozov}
\email{i.a.morozov@inp.nsk.su}
\affiliation{Budker Institute of Nuclear Physics,
11 Acad. Lavrentieva Pr., Novosibirsk, 630090 Russia}
\date{\today}

\begin{abstract}
In this article we consider two dynamical systems: the McMillan
sextupole and octupole integrable mappings originally introduced
by Edwin McMillan;
the second one is also known as canonical McMillan map.
Both of them are simplest symmetric McMillan maps with only one
intrinsic parameter, the trace of the Jacobian at the fixed point.
While these dynamical systems have numerous of applications and
are used in many areas of math and physics, some of their dynamical
properties have not been described yet.
We fulfill the gap and provide complete description of all stable
trajectories including parametrization of invariant curves,
Pioncar\'e rotation numbers and canonical action-angle variables.

In the second part we relate these maps with general chaotic map
in McMillan-Turaev form.
We show that McMillan sextupole and octupole mappings are first
order approximations of dynamics around the fixed point, in a
similar way as linear map and quadratic invariant (Courant-Snyder
invariant in accelerator physics) is the zeroth order approximation
(known as linearization).
Finally we suggest the new formalism of nonlinear Twiss parameters
which incorporate dependence of rotation number as a function of
amplitude, in contrast to e.g. betatron phase advance used in
accelerator physics which is independent of amplitude.
Specifically in application to accelerator physics this new formalism
is capable of predicting dynamical aperture around 1-st, 2-nd, 3-rd
and 4-th order resonances for flat beams, which is critical for
beam injection/extraction.
\end{abstract}

\maketitle
\tableofcontents

\newpage
\section{Introduction}

A generic dynamical system typically endures the presence of chaos.
The motion is chaotic when there is an insufficient amount of
integrals of motion, also known as invariants or conserved
quantities.
Such systems are difficult to analyze and known to be intractable
problems, i.e. any practical solution might require too many
computational resources to be useful due to its high sensitivity to
initial conditions, rounding errors and external perturbations.

On the other hand, when the system possesses enough first integrals,
it is known as {\it integrable} with quite simple and predictable
dynamics.
Historically, many completely integrable problems studied in physics
played an important role in further understanding more realistic and
complicated scenarios.
For example:
(i) a multi-dimensional harmonic oscillator works as a good
approximation of small amplitude oscillations in a general nonlinear
system,
(ii) the Kepler problem gives a general understanding of, let's say,
Earth's orbit while the entire solar system consists of many objects
(many-body problem) and is highly chaotic, and
(iii) planetary motion about two fixed centers has been used in the
calculation of satellite trajectories in the gravitational field of
the
Earth~\cite{alexeev1965generalized,marchal1966calcul,marchal1986quasi},
the acceleration of electrons in atomic
collisions~\cite{jakas1995trapping,jakas1996production}, and, the
calculation of the energy levels of the
${\mathrm{H}_2}^+$~\cite{strand1979semiclassical}.

In this article, we focus on another famous example known as
McMillan integrable map~\cite{mcmillan1971problem}.
Originally, it was proposed by Edwin McMillan as a simple
accelerator lattice with one degree of freedom, consisting of
linear optics (corresponding to simple linear transformation in
phase space) and a special thin nonlinear lens (nonlinear vertical
shear transformation).
As later demonstrated by Suris~\cite{suris1989integrable}, McMillan
mapping is also one of the very few possible integrable symplectic
transformations of the plane with analytic integral of motion;
for the map in the form~(\ref{math:McMmap}) invariant has to be
either regular, exponential or trigonometric polynomial of degree
two in coordinate and momentum.

Here we demonstrate that in fact, McMillan map is not only a model
lattice, but can be related to the general planar symplectic map
with chaotic dynamics as its first order approximation
(Section~\ref{sec:PertTh});
the zeroth order considers local stability of an equilibrium point,
known as a linearization of the map.
This fact helps to ``integrate-out'' additional nonlinear features of
small amplitude oscillations, including dependence of frequency on
action variable.
As a consequence, we suggest the new concept of nonlinear Twiss
parameters for accelerator physics which is the natural extension
to the existing Courant-Snyder formalism (Section~\ref{sec:Twiss}).
The new ``language'' allows not only an understanding of
amplitude-dependent shift of the betatron tune, but a prediction
of dynamical aperture around isolated low order resonances,
representing a nonlinear integrable model for slow resonant
extraction of a flat beam (Section~\ref{sec:AccExample}).

Finally, in first part of the article, we fill the gap in analytical
results by providing a canonical transformation to action-angle
variables (Section~\ref{sec:Dynamics}) and complimentary bifurcation
diagrams for McMillan sextupole and octupole mappings
(Section~\ref{sec:Bifurcations});
these are the simplest symmetric McMillan maps with only one
intrinsic parameter and quadratic (sextupole) or cubic (octupole)
nonlinearities.
Octupole, and even more general asymmetric McMillan maps, have
been studied in great detail, especially by Iatrou and
Roberts~\cite{IR2002II} who provided parametrization of individual
curves.
However, explicit expressions for rotation number as a function of
canonical action integral have been missing until now.
This knowledge is critical in understanding underlying dynamics
and design of integrable accelerator lattices with desirable
properties.
Additionally, these analytical results can be applied to more
realistic integrable accelerator lattices based on McMillan map,
e.g. they describe a particle with zero angular momentum in the
integrable (in 4D phase space) accelerator ring with axially
symmetric McMillan electron lens~\cite{cathey2021calculations}.

\newpage
\section{McMillan sextupole and octupole mappings}

In this section we define two dynamical systems: {\it McMillan
sextupole} and {\it McMillan octupole} integrable mappings.
Both of them are in {\it McMillan-Turaev form} (or MT form for
short), defined as:
\begin{equation}
\label{math:McMmap}
\begin{array}{ll}
    \mathcal{M}_\mathrm{MT}:    & q' = p,       \\[0.25cm]
                                & p' =-q + f(p).
\end{array}
\end{equation}

The sextupole map has a force function
\begin{equation}
\label{math:fsexExp}
    f_\text{sex}(p) =-p\,\frac{p+2\,\epsilon}{p+\Gamma} =
   -\frac{2\,\epsilon}{\Gamma}\,p
   - \frac{\Gamma-2\,\epsilon}{\Gamma^2}\,p^2
   + \frac{\Gamma-2\,\epsilon}{\Gamma^3}\,p^3
   +\mathcal{O}(p^4).
\end{equation}
The first nonlinear term in the power series is quadratic,
corresponding to the thin sextupole lens.
When $f(p) = a\,p + b\,p^2$, the map is non-integrable and is known
as the {\it area-preserving H\'enon quadratic map},
\cite{henon1969numerical}.
In Section~\ref{sec:Remarks} we provide the relation between general
map in MT form (including H\'enon and Chirikov maps) and
McMillan sextupole and octupole mappings as its approximation.

The octupole map is sometimes referred to as {\it canonical
McMillan Map} (see \cite{IR2002II}) with
\begin{equation}
\label{math:foctExp}
    f_\text{oct}(p) =-\frac{2\,\epsilon\,p}{p^2+\Gamma} =
   -\frac{2\,\epsilon}{\Gamma}\,p
   +\frac{2\,\epsilon}{\Gamma^2}\,p^3
   -\frac{2\,\epsilon}{\Gamma^3}\,p^5
   +\mathcal{O}(p^7).
\end{equation}
Now the first nonlinear term is cubic, corresponding to the thin
octupole lens.
We suggest these names in order to reflect
the dynamical nature of systems.
This integrable map approximates general mappings of the
form~(\ref{math:McMmap}) when $f(p)$ is an odd function, or more
generally when $\pd_{pp} f(p) = 0$.

\subsection{Symmetries of invariants of motion}

The invariant of motion for the sextupole map in matrix form is
\[
\begin{bmatrix}
p^2 \\ p  \\ 1
\end{bmatrix}^\mathrm{T}
\left(
\begin{bmatrix}
0		& 1			    & \Gamma	\\
1		& 2\,\epsilon	& 0			\\
\Gamma	& 0			    &-\K
\end{bmatrix}
\cdot
\begin{bmatrix}
q^2 \\ q  \\ 1
\end{bmatrix}
\right) =
p^2\,q + p\,q^2 + \Gamma\,(p^2 + q^2) + 2\,\epsilon\,p\,q - \K \equiv 0,
\]
or explicitly
\[
   \K_\text{sex}(p,q) = p^2\,q + p\,q^2 + \Gamma\,(p^2 + q^2) + 2\,\epsilon\,p\,q.
\]
The following propositions are straightforward to verify.
\begin{proposition}
\label{prop:Sex1}
Simultaneous change of map parameters ($\varepsilon \neq 0$)
\[
    \Gamma \rightarrow \varepsilon\,\Gamma,
    \qquad\qquad
    \epsilon \rightarrow \varepsilon\,\epsilon,
\]
along with the scaling transformations
\[
(p,q) \rightarrow \varepsilon\,(p,q)
\qquad\mathrm{and}\qquad
\K \rightarrow \varepsilon^3\,\K
\]
leaves the form of the map and biquadratic invariant.
\end{proposition}
\begin{proposition}
\label{prop:Sex2}
Simultaneous change of map parameters
\[
    \Gamma \rightarrow \frac{\Gamma-2\,\epsilon}{3},
    \qquad\qquad
    \epsilon \rightarrow-\frac{4\,\Gamma + \epsilon}{3},
\]
along with the translation transformations
\[
(p,q) \rightarrow (p,q) + \frac{2}{3}\,(\Gamma+\epsilon)
\qquad\mathrm{and}\qquad
\K \rightarrow \K - \left[ \frac{2}{3}\,(\Gamma+\epsilon) \right]^3
\]
leaves the form of the map and biquadratic invariant.
\end{proposition}

There are two consequences from the first proposition.
First, there is only one intrinsic parameter --- the ratio
of $\epsilon$ and $\Gamma$;
in a space of map parameters $(\Gamma,\epsilon)$, all dynamical
systems on a ray $\epsilon/\Gamma = \const$ that starts at the
origin, are similar up to the scaling transformation.
In fact, this parameter is the trace of Jacobian evaluated at the
fixed point at the origin, $\zeta^{(1-1)}=(0,0)$,
\[
    \Tr\,\J_\text{sex}(\zeta^{(1-1)}) =-\frac{2\,\epsilon}{\Gamma} = a.
\]
Second, Proposition~\ref{prop:Sex1} tells us that if we know
the dynamics on a ray $\epsilon/\Gamma = \const$, the dynamics for
the systems on the opposite ray, $\Gamma\rightarrow-\Gamma$ and
$\epsilon\rightarrow-\epsilon$, is given by the rotation of the
phase space by an angle of $\pi$ and inversion of $\K$,
(case $\varepsilon =-1$).

The second proposition shows that there is a discrete symmetry with
respect to the choice of origin; translation
\[
    (p,q) \rightarrow (p,q) - \zeta^{(1-2)}
\]
shifts the origin to the second fixed point located on the main
diagonal
\[
    \zeta^{(1-2)} = -\frac{2}{3}\,(1,1)\,(\Gamma+\epsilon).
\]
The transformation of parameters
\[
\begin{bmatrix}
\Gamma \\[0.2cm] \epsilon
\end{bmatrix} \rightarrow
\begin{bmatrix}
 \frac{1}{3} &-\frac{2}{3} \\[0.2cm]
-\frac{4}{3} &-\frac{1}{3}
\end{bmatrix}
\begin{bmatrix}
\Gamma \\[0.2cm] \epsilon
\end{bmatrix},
\]
is an inversion ($\det = -1$) with two eigenvectors: the eigenvector
along the line $\epsilon = -\Gamma$ with unit eigenvalue
\[
    v_1 = (-1,1)/\sqrt{2},
    \qquad
    \lambda_1 = 1,
\]
and another eigenvector with minus unit egenvalue and along the line
$\epsilon = 2\,\Gamma$
\[
    v_2 = (1,2)/\sqrt{5},
    \qquad
    \lambda_2 =-1.
\]

\begin{figure}[h!]
    \centering
    \includegraphics[width=0.3\linewidth]{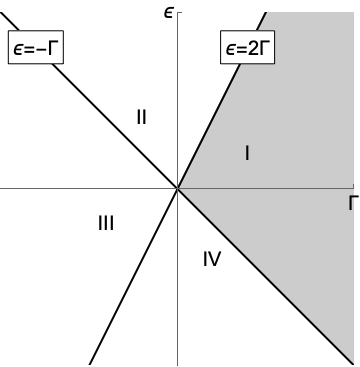}
    \caption{\label{fig:SexSymm}
    Symmetry lines on the $(\Gamma,\epsilon)$-plane for McMillan
    sextupole map.
    Dynamics in sectors I and III (II and IV) are related through
    the Proposition~\ref{prop:Sex1}.
    Dynamics in sectors I and IV (II and III) are related through
    the Proposition~\ref{prop:Sex2}.
    Filled region shows the area of considered parameters.
    }
\end{figure}

Both eigenvectors are related to symmetries between fixed points
$\zeta^{(1-1,2)}$.
For all dynamical systems along the line $\epsilon=-\Gamma$, fixed
points $\zeta^{(1-1,2)}$ undergo transcritical bifurcation and
degenerate to a single point.
When $\epsilon = 2\,\Gamma$, the phase space portrait of such
dynamical systems has mirror symmetry with respect to line
$p=-q-\frac{1}{2}$, and, $\zeta^{(1-1,2)}$ are mirror images
of each other.
Without loss of generality, as a result of Propositions
\ref{prop:Sex1}--\ref{prop:Sex2}, we will consider the dynamics
only for the area of parameters (sector I in Fig.~\ref{fig:SexSymm})
\[
    (\epsilon >-\Gamma)
    \,\cap\,
    (\epsilon <2\,\Gamma)
    \qquad\mathrm{or}\qquad
    a \in [-4;2],\quad \Gamma \geq 0.
\]

The invariant of motion for the octupole map is
\[
\begin{bmatrix}
p^2 \\ p  \\ 1
\end{bmatrix}^\mathrm{T}
\left(
\begin{bmatrix}
1		& 0			    & \Gamma	\\
0		& 2\,\epsilon   & 0			\\
\Gamma	& 0			    &-\K
\end{bmatrix}\cdot
\begin{bmatrix}
q^2 \\ q  \\ 1
\end{bmatrix}
\right) =
p^2 q^2 + \Gamma\,(p^2 + q^2) + 2\,\epsilon\,p\,q - \K \equiv 0,
\]
or
\[
\K_\text{oct}(p,q) = p^2 q^2 + \Gamma\,(p^2 + q^2) + 2\,\epsilon\,p\,q,
\]
with the following propositions
\begin{proposition}
Simultaneous change of map parameters ($\varepsilon > 0$)
\[
    \Gamma \rightarrow \varepsilon\,\Gamma,
    \qquad\qquad
    \epsilon \rightarrow \varepsilon\,\epsilon,
\]
along with the scaling transformations
\[
(p,q) \rightarrow \sqrt{\varepsilon}\,(p,q)
\qquad\mathrm{and}\qquad
\K \rightarrow \varepsilon^2\,\K
\]
leaves the form of the map and biquadratic invariant.
\end{proposition}

\begin{proposition}
Change of the map parameter
\[
    \epsilon \rightarrow -\epsilon,
\]
along with the reflection with respect to $p=0$ (or $q=0$),
\begin{equation}
\label{math:trans}
\begin{array}{l}
q' = q     \\[0.25cm]
p' =-p
\end{array}
\qquad\left(\text{or}\qquad
\begin{array}{l}
q' =-q     \\[0.25cm]
p' = p
\end{array}\right)
\end{equation}
leaves the biquadratic invariant.
\end{proposition}

As a result of the first proposition, similar to the sextupole case,
the dynamics on ray $\epsilon/\Gamma = \const$ with initial point
at the origin is identical up to the scaling transformation.
While the system still has one intrinsic parameter
$a=-2\,\epsilon/\Gamma$, in contrast to the sextupole case, we have
two different forms of invariants for $\Gamma \lessgtr 0$, since
$\varepsilon$ is strictly positive.

The second proposition is less powerful, since it only guarantees
the conservation of the form of the invariant but not the map.
In fact, it can be accompanied by another proposition:
\begin{proposition}
Change of the map parameter
\[
    \epsilon \rightarrow -\epsilon,
\]
influences the dynamics of stable trajectory on a given energy
level, $\K = \const$, as follows:
\begin{itemize}
    \item The invariant curves $\K(q,p) \rightarrow \K(q,-p)=\K(-q,p)$.
    \item Symmetric fixed points transform to 2-cycle, and vice versa
          2-cycle transforms to a pair of symmetric fixed points.
    \item Dynamics around fixed point at the origin transforms according
          to
    \[
    \begin{array}{l}
    J \rightarrow J,     \\[0.25cm]
    \nu \rightarrow \frac{1}{2} - \nu,
    \end{array}
    \]
    where $J$ is the action and $\nu$ is the rotation number.
\end{itemize}
\end{proposition}

$\bigtriangleup$
\begin{itemize}
\item Obvious by direct substitution.
\item In the next section it is shown explicitly by obtaining all fixed
      points and 2-cycles, but here we will provide some geometrical
      considerations.
      Due to the form of the map, all fixed points belong to the
      intersection of main diagonal $p=q$ and second symmetry line
      $p=f_\mathrm{oct}(q)/2$.
      All 2-cycles belong to the intersection of the second symmetry line
      with its inverse, i.e. $q=f_\mathrm{oct}(p)/2$, and
      since $f_\mathrm{oct}$ is an odd function of its argument, the only
      solutions allowed belong to anti-diagonal $p=-q$.
      After the transformation $(p,q)\rightarrow(-p,q)$, corresponding
      critical points of the invariant from the main diagonal mapped to
      the anti-diagonal and vice versa.
      Additionally, it automatically follows from the next part of
      the proposition.
\item Since after transformation~(\ref{math:trans}) the area under
      the closed curve is preserved, it follows $J \rightarrow J$.
      The action can be seen as $J = 2\,J' + 2\,J''$, where $J'$ is
      an area over $2\pi$ in II and IV, and, $J''$ is in I and III
      quadrants of $(p,q)$-plane.
      Using Danilov theorem
      \cite{zolkin2017rotation,nagaitsev2020betatron},
      $\nu = \dd J'/\dd J$ and that after the transformation
      (\ref{math:trans}) $J' \leftrightarrow J''$ one has:
        \[
            \nu \rightarrow \frac{\dd J''}{\dd J} =
            \frac{\dd\left( \frac{J}{2}-J' \right)}{\dd J} = 
            \frac{1}{2} - \nu.
        \]
\end{itemize}
$\bigtriangledown$

Another way to see that both mappings have only one natural
parameter is to notice that both force functions have singularities:
$p =-\Gamma$ for the sextupole and $p = \pm\sqrt{\Gamma}$ for the
octupole maps.
Performing the non-dimensionalization by measuring $q$ and $p$ in
units of $\Gamma$ or $\sqrt{|\Gamma|}$ for the sextupole and octupole
maps respectively, one obtains
\[
\K_\text{sex}(q,p)/\Gamma^3 = \Sigma\,\Pi + \cs
\]
and
\[
\K_\text{oct}(q,p)/|\Gamma|^2 = \Pi^2 \pm \cs,
\qquad\text{for }
\Gamma \gtrless 0,
\]
where $\Sigma = p + q$ and $\Pi = p\,q$ are symmetric combinations
of dynamic variables and
\[
    \cs \equiv p^2 + \frac{2\,\epsilon}{\Gamma}\,p\,q + q^2 =
        p^2 - a\,p\,q + q^2 =
        \Sigma^2 -(a+2)\,\Pi
\]
is the {\it Courant-Snyder term}.
Now nonlinear terms appear with unit coefficients and the only
parameter left is the trace of Jacobian at the origin $a$.

\newpage
\section{\label{sec:Bifurcations}Stability of fixed points and 2-cycles. \\
         Bifurcation diagrams.}

{\it Fixed point} ($n=1$) and {\it $n$-cycle} of the map $\mathcal{M}$,
$\zeta^{(n)}\equiv(q,p)$, are defined by $\mathcal{M}^n (q,p) = (q,p)$.
Stability of $n$-cycle is given by the trace of Jacobian,
$\left|
    \Tr\,\J[\mathcal{M}^n (\zeta^{(n)})]
\right| < 2$.
Analysis of isolated $n$-cycles of the map plays a critical role in
understanding its bifurcations.
Iatrou and Roberts \cite{IR2002II} proved that for an integrable
map with a smooth integral, isolated critical points of the integral
belong to (isolated) cycles of the map and the points of isolated
cycles of the map are (isolated) critical points of the integral.
Applying it to a general asymmetric McMillan map, one can verify that
invariant of motion can have up to 5 isolated critical points.
Without a loss of generality, we assume that one of them corresponds
to a fixed point at the origin.
Four other critical points have to be fixed points as well, or,
two of them can form an isolated 2-cycle.
In an exceptional case these fixed points and 2-cycles can appear
on the same level of invariant and then form 3- or 4-cycles;
then, the map degenerates to a linear and, in fact, all orbits (except
fixed point at the origin) become periodic with corresponding
periods 3 and 4.
Below we will consider bifurcations of the sextupole and octupole
McMillan mappings.

For the mapping in McMillan form, fixed points belong to the main
diagonal $p=q$ and are given by the intersection of two symmetry
lines
\[
    p=q
    \qquad\mathrm{and}\qquad
    p = f(q)/2.
\]
2-cycles are given by the intersection of the second symmetry line
with its inverse
\[
    p = f(q)/2
    \qquad\mathrm{and}\qquad
    q = f(p)/2.
\]
If $f(q)$ is an odd function, then fixed points have to appear in
symmetric pairs with respect to the origin and 2-cycles belong to
intersection of anti-diagonal $p=-q$ with $p = f(q)/2$.

\vspace{0.5cm}

\subsection{Sextupole map}

The sextupole map has two fixed points.
The first one is at the origin
\[
    \zeta^{(1-1)} = (0,0),
\]
and an additional fixed point on the main diagonal is
\[
    \zeta^{(1-2)} = -\frac{2}{3}\,(\Gamma+\epsilon)\,(1,1) =
    \frac{a-a_0}{3}\,(1,1)\,\Gamma.
\]
Both fixed points are defined for any values of parameters on
$(\Gamma,\epsilon)$-plane.
In addition, there is a 2-cycle
\[
\begin{array}{l}
    \zeta^{(2)} = (\Gamma-\epsilon)\,(1,1) \pm
    \sqrt{(\Gamma-\epsilon)(5\,\Gamma-\epsilon)}\,(1,-1)    \\[0.25cm]
\ds = \frac{1}{2}\,\left[
    (a-a_{1/2})\,(1,1) \pm
    \sqrt{ (a-a_{1/2}) (a-a^*_{1/2})}\,(1,-1)
    \right]\,\Gamma,
\end{array}
\]
defined only in a sector $(\Gamma-\epsilon)(5\,\Gamma-\epsilon)>0$.
Stability analysis shows that 2-cycle is always unstable when
$\zeta^{(2)}$ is real, and is stable when
$\zeta^{(2)}\in \mathbb{C}^2$.

\begin{figure}[h!]
    \centering
    \includegraphics[width=0.89\linewidth]{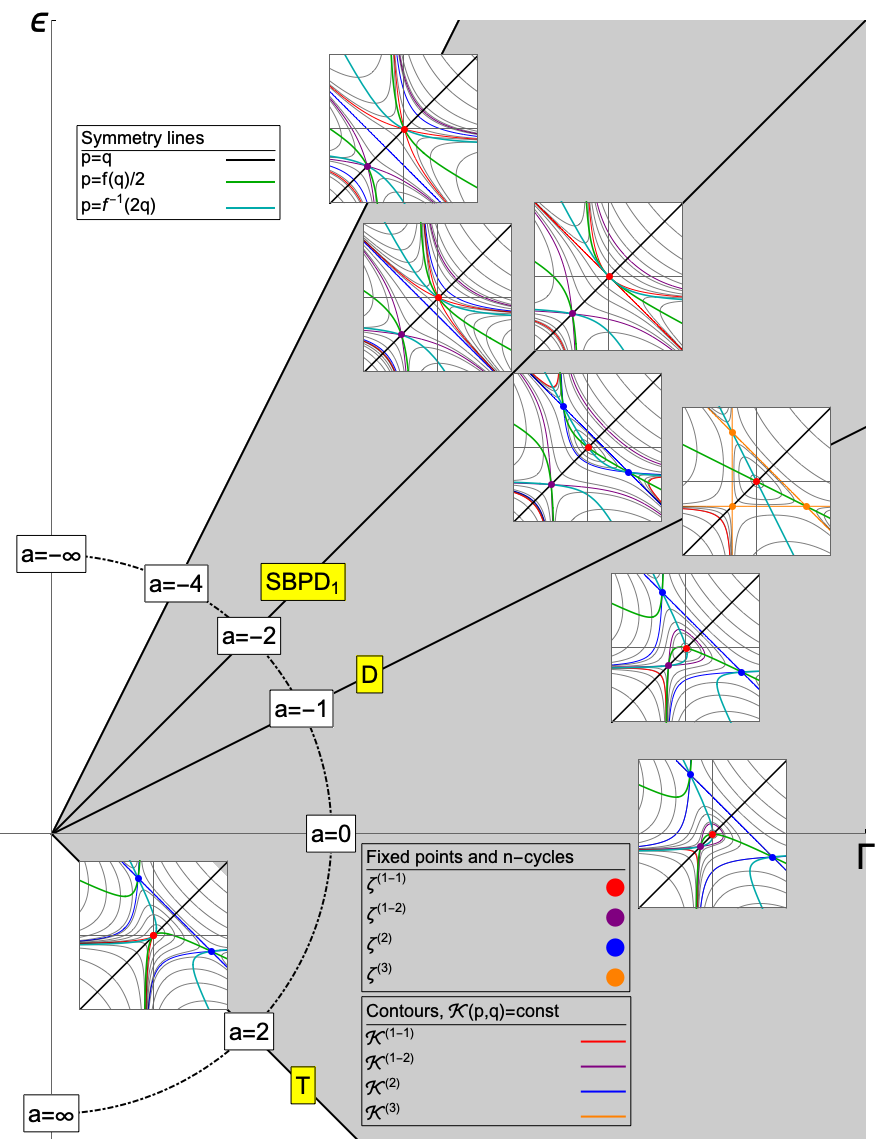}
    \caption{\label{fig:BifDSbig}
    Bifurcation diagram for the McMillan sextupole map.
    $(\Gamma,\epsilon)$-plane with rays of constant values of
    $a=-2\,\epsilon/\Gamma$.
    Sub-critical period doubling and transcritical bifurcations
    are marked with SBPD and T respectively.
    Label D corresponds to the system degeneracy.
    For each bifurcation (or in between) we provide characteristic
    contour plot of constant level sets of invariant.
    On each contour plot, fixed points, 2- and 3-cycles, as well
    as their corresponding level sets, are shown in colors
    according to the legend;
    other level sets are shown in black.
    First symmetry line $p=q$, second symmetry line $p=f(q)/2$
    and its inverse $q=f(p)/2$ are shown in thick black, green
    and cyan respectively.
    }
\end{figure}

A sophisticated version of the bifurcation diagram including level sets
of invariant is presented in Fig~\ref{fig:BifDSbig}.
The plot shows $(\Gamma,\epsilon)$-plane (for $\Gamma > 0$) with
additional scale for $a=-2\,\epsilon/\Gamma$ (added on the dash-dotted
circle).
Each ray on the diagram corresponds to the constant value of $a$
and contour plots of invariant $\K(p,q)$ are included, illustrating
possible dynamical scenarios and bifurcations.

\begin{figure}[ht]
    \centering
    \includegraphics[width=0.74\linewidth]{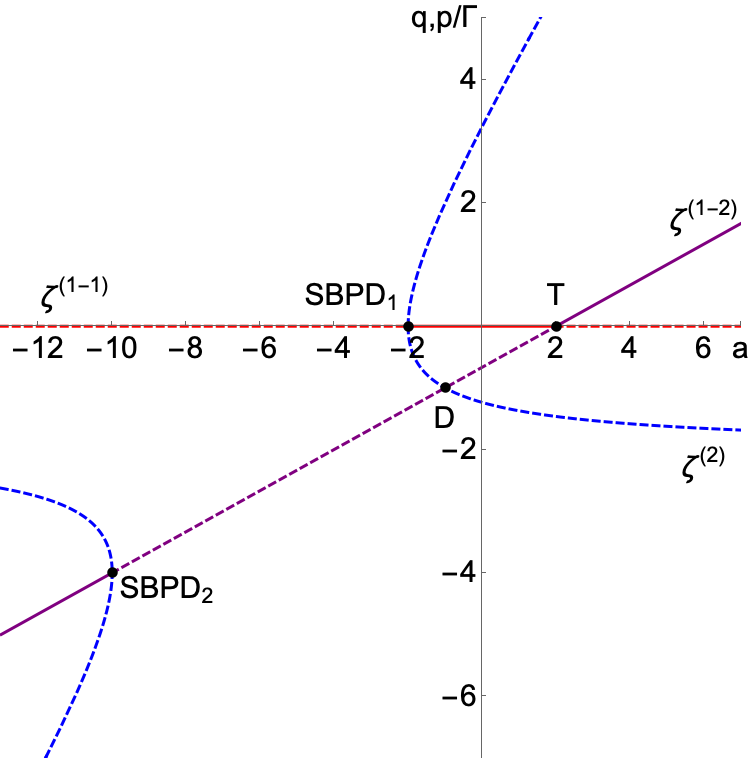}
    \caption{\label{fig:BifDSsmall}
    Bifurcation diagram of fixed points and 2-cycles for the McMillan
    sextupole map.
    Plot shows normalized coordinates of 1,2-cycles as a function of
    parameter $a=-2\,\epsilon/\Gamma$.
    When fixed point or 2-cycle is stable (unstable) then it is shown as
    solid (dashed).
    Sub-critical period doubling and transcritical bifurcations are
    marked with SBPD and T respectively.
    Label D corresponds to the value of $a=-1$ and represents system
    degeneracy.
    }
\end{figure}

Stability and bifurcations of fixed points (2-cycles) are summarized
in Fig.~\ref{fig:BifDSsmall}.
The fixed point at the origin is stable for $a_{1/2} = -2 < a < a_0 = 2$
(lower index for particular values of $a_\nu$ is the corresponding
rotation number $\nu = \arccos(a/2)/(2\pi)$).
When $a=a_0$ it experiences transcritical bifurcation and exchange
stability with $\zeta^{(1-2)}$.
When $a=a_{1/2}$, $\zeta^{(1-1)}$ undergoes sub-critical period
doubling bifurcation, the
second fixed point also goes through sub-critical period doubling
at $a=a^*_{1/2}=-10$.
Finally, for $a=a_{1/3}=-1$ the second fixed point and 2-cycle approach
the same energy level and the map degenerates to a linear with $f(p) =-p$
and all orbits around the origin have period equal to 3.

\subsection{Octupole map}

Similar to the sextupole case, the octupole map has a fixed point
at the origin
\[
    \zeta^{(1-1)} = (0,0),
\]
which is again defined $\forall\,(\Gamma,\epsilon)\in \mathbb{R}^2$
and stable for $|\epsilon|<|\Gamma|$ (or $a_{1/2} < a < a_{0}$).
In contrast, it has two more fixed points appearing
in a symmetric pair on the same energy level of the invariant
\begin{equation}
\label{math:zeta1-23}
    \zeta^{(1-2,3)} = \pm\sqrt{-(\Gamma+\epsilon)}\,(1,1)
    = \pm\sqrt{\pm\frac{a-a_{0}}{2}}\,(1,1)\sqrt{|\Gamma|}.
\end{equation}
Note that the $\pm$-sign under the square root is for the cases
$\Gamma \gtrless 0$
(for both Eqs.~(\ref{math:zeta1-23},\ref{math:zeta2})).
Also note that in comparison to the sextupole map, the ``natural''
unit of distance is $\sqrt{|\Gamma|}$ instead of $\Gamma$.
Symmetric fixed points are defined only for $\epsilon<-\Gamma$
(or $a \gtrless a_0$ for $\Gamma \gtrless 0$).
A 2-cycle is defined for
$\epsilon>\Gamma$ (or $a \lessgtr a_{1/2}$ for $\Gamma \gtrless 0$):
\begin{equation}
\label{math:zeta2}
    \zeta^{(2)} =\pm\sqrt{-(\Gamma-\epsilon)}\,(1,-1)
    = \pm\sqrt{\mp\frac{a-a_{1/2}}{2}}\,(1,-1)\sqrt{|\Gamma|}.
\end{equation}

\begin{figure}[h!]
    \centering
    \includegraphics[width=\linewidth]{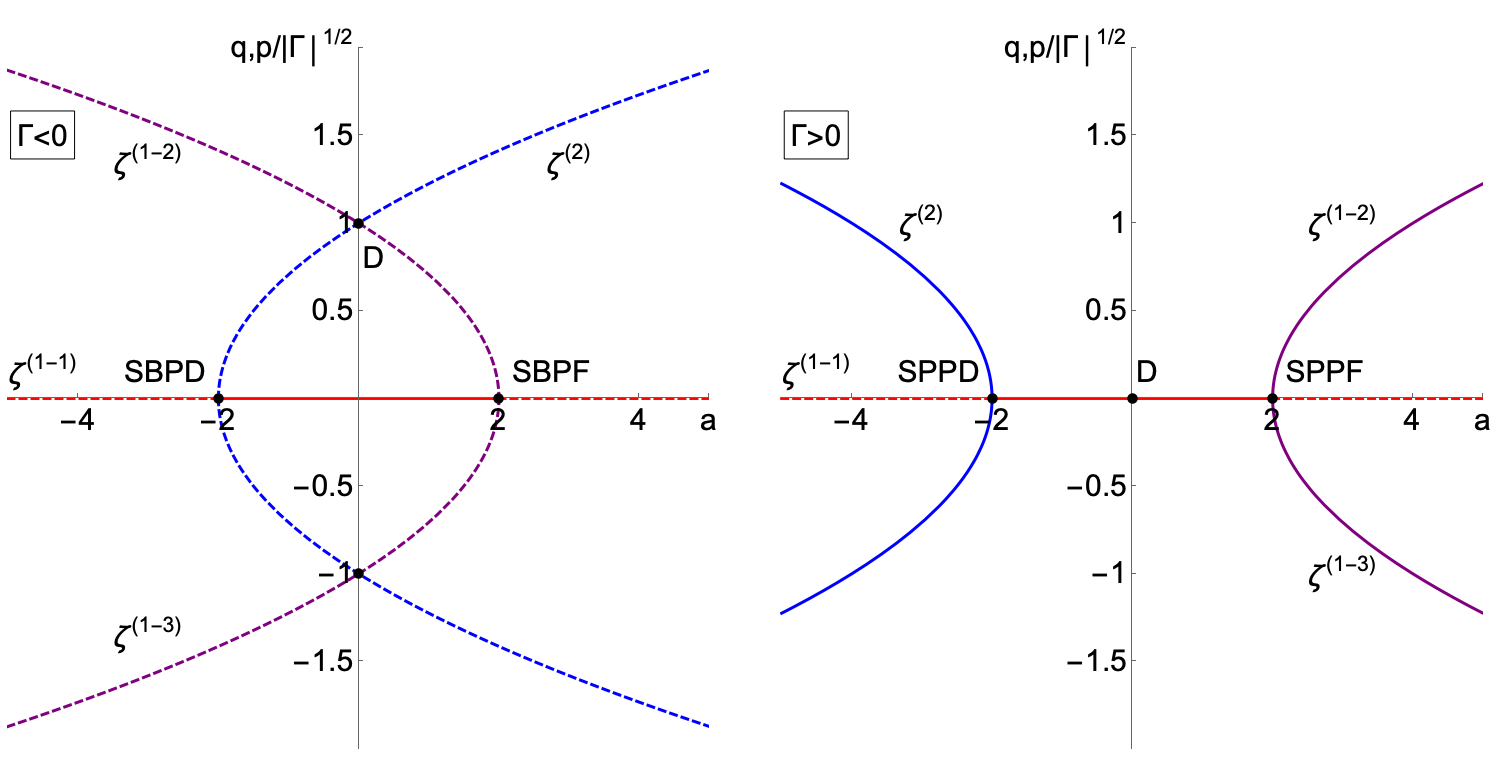}
    \caption{\label{fig:BifDOsmall}
    Bifurcation diagram of invariant points and 2-cycles for the McMillan
    octupole map.
    Left and right plots show normalized coordinates of 1,2-cycles as
    a function of parameter $a=-2\,\epsilon/\Gamma$ for $\Gamma < 0$
    and $\Gamma>0$ respectively.
    When fixed point or 2-cycle is stable (unstable) then it is shown as solid
    (dashed).
    Pitchfork and period doubling bifurcations are marked with PF and PD
    (SB for sub-critical and SP for super-critical).
    Label D corresponds to the value of $a=0$ and represents system
    degeneracy.
    }
\end{figure}

Stability and bifurcations plots for cases $\Gamma \gtrless 0$ are
presented in Fig.~\ref{fig:BifDOsmall}.
When $\Gamma < 0$ at $a_{1/2,0}$ the fixed point undergoes sub-critical
period doubling or pitchfork bifurcations respectively.
Fixed points $\zeta^{(1-2,3)}$ and 2-cycle $\zeta^{(2)}$ are always
unstable and the only region with stable trajectories is $a_{1/2}<a<a_0$.
For $\Gamma > 0$ at $a_{1/2,0}$ the fixed point undergoes super-critical
period doubling or pitchfork bifurcations.
When $\zeta^{(1-1)}$ is stable, $a_{1/2}<a<a_0$, there are no additional
fixed points or 2-cycles and all trajectories are stable.
When $|a|>2$, all trajectories are globally stable again, but the point
in the origin is locally unstable and becomes a center of lemniscate.
If $a>a_0$, trajectories inside the figure-eight curve round the centers of
``eyes,'' while when $a<a_{1/2}$ the orbit jumps from ``eye'' to ``eye'' since the centers become 2-cycles.

An additional bifurcation diagram with level sets of invariant is
show below in Fig~\ref{fig:BifDObig}.
Similar to the sextupole case, we use $(\Gamma,\epsilon)$-plane
but now with two additional scales for $a$ on each semi-plane
$\Gamma \gtrless 0$.
Each ray on a diagram corresponds to the constant value of $a$
and contour plots of invariant $\K(p,q)$ are included, illustrating
possible dynamical scenarios and bifurcations.

\begin{figure}[p]
    \centering
    \includegraphics[width=\linewidth]{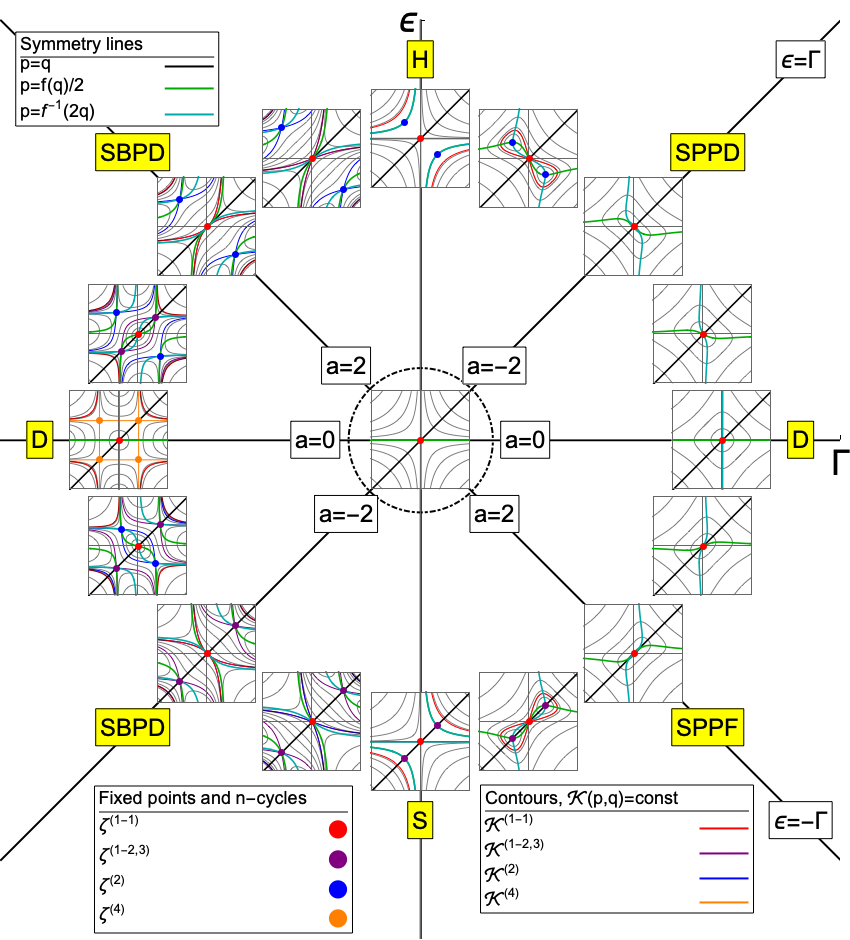}
    \caption{\label{fig:BifDObig}
    Bifurcation diagram for the McMillan octupole map.
    $(\Gamma,\epsilon)$-plane with rays of constant values of
    $a=-2\,\epsilon/\Gamma$
    (one scale for each semi-plane $\Gamma \gtrless 0$).
    Pitchfork and period doubling bifurcations are marked with PF and PD
    (SB for sub-critical and SP for super-critical).
    Label D corresponds to the system degeneracy.
    For each bifurcation (or in between) we provide characteristic
    contour plot of constant level sets of invariant.
    On each contour plot, fixed points, 2- and 4-cycles as well
    as their corresponding level sets are shown in colors
    according to the legend;
    other level sets are shown in black.
    First symmetry line $p=q$, second symmetry line $p=f(q)/2$
    and its inverse $q=f(p)/2$ are shown in thick black, green
    and cyan respectively.
    }
\end{figure}

\newpage
\section{\label{sec:Dynamics}Dynamics of sextupole and octupole \\ McMillan mappings}

\subsection{Sextupole map}

\subsubsection{Parametrization of individual curve}

Here we will illustrate the general method of obtaining a
parametrization of an individual curve using the sextupole
map as an example.
We limit ourselves to cases with stable trajectories around
the stationary point at the origin, considering only $a_{1/2}<a<a_0$.
We start with a formal Hamiltonian system
\[
    \h[p,q;t] \equiv \K(p,q) =
    p^2q + p\,q^2 + p^2 -a\,p\,q + q^2
\]
which satisfies system of Hamilton's equations
\[
\begin{array}{r}
\ds\qquad  \frac{\dd q}{\dd t} = \frac{\pd\h}{\pd p} =
    q^2 - a\,q + 2\,p + 2\,p\,q,       \\[0.25cm]
\ds-       \frac{\dd p}{\dd t} = \frac{\pd\h}{\pd q} =
    p^2 - a\,p + 2\,q + 2\,p\,q.
\end{array}
\]
Since Hamiltonian function does not change along the trajectory,
we can solve for $p$ from $\h = \mathrm{const}$:
\[
\begin{array}{rl}
p\!\!\!& \ds = \frac{q\,(a-q)\pm
        \sqrt{  q^4 -
                2\,(2 + a)\,q^3 + 
                (a^2-4)\,q^2 +
                4\,\h\,q +
                4\,\h
        }
    }{2\,(q+1)}                                \\[0.25cm]
    & \ds = \frac{1}{2}\,\left[
    f_\text{sex}(q) \pm \frac{\sqrt{\mathcal{P}(q)}}{q+1}
\right],
\end{array}
\]
where
\[
\mathcal{P}(q) =
    q^4 - 2\,(a-a_{1/2})\,q^3 + (a-a_{0})(a-a_{1/2})\,q^2 +
    4\,\h\,q + 4\,\h.    
\]
Substitution into the first Hamilton's equation gives
\begin{equation}
\label{math:intRAW}
    \dd\,t = \pm\frac{\dd\,q}{\sqrt{\mathcal{P}(q)}}.
\end{equation}
When the fixed point at the origin is stable ($a_{1/2}<a<a_0$),
polynomial $\mathcal{P}(q)$ has four distinct real roots such that
\[
    q_1 < q_2 \leq q \leq q_3 < q_4
\]
and $\h > 0$ for all closed invariant curves.
Roots $q_i$ satisfies
\[
    \sum_{i=1}^{4}\frac{1}{q_i} = -1
\]
and using Vieta's formulas for the quartic $\mathcal{P}(q)$,
one can express map parameter and energy level using $q_i$:
\[
    a = -2 + \sum_{i=1}^{4}q_i/2
    \qquad\mathrm{and}\qquad
    \h = \frac{1}{4}\prod_{i=1}^{4} q_i.
\]

Factorizing the polynomial, $\mathcal{P}(q) = (q_4-q)(q_3-q)(q-q_2)(q-q_1)$,
we rewrite Eq.~(\ref{math:intRAW}) as an integral
\[
\int_0^t \dd\,t =
    \pm \int_{q_0}^q\frac{\dd\,q}{\sqrt{(q_4-q)(q_3-q)(q-q_2)(q-q_1)}}.
\]
This is the standard integral which can be expressed in terms of
any of three principal Jacobian elliptic functions
\[
\begin{array}{rl}
q(t)\!\!\!\!
    &\ds = \frac{q_2 - q_1\,\frac{q_3-q_2}{q_3-q_1}\,\sn_1^2}
            {1-\frac{q_3-q_2}{q_3-q_1}\,\sn_1^2}
      = \frac{q_3 - q_4\,\frac{q_3-q_2}{q_4-q_2}\,\sn_2^2}
            {1-\frac{q_3-q_2}{q_4-q_2}\,\sn_2^2}              \\[0.5cm]
    &\ds = \frac{q_3 + q_1\,\frac{q_3-q_2}{q_2-q_1}\,\cn_1^2}
            {1+\frac{q_3-q_2}{q_2-q_1}\,\cn_1^2}
      = \frac{q_2 + q_4\,\frac{q_3-q_2}{q_4-q_3}\,\cn_2^2}
            {1+\frac{q_3-q_2}{q_4-q_3}\,\cn_2^2}              \\[0.5cm]
    &\ds = \frac{q_2 - q_1\,\frac{q_4-q_2}{q_4-q_1}\,(1-\dn_1^2)}
            {1-\frac{q_4-q_2}{q_4-q_1}(1-\dn_1^2)}
      = \frac{q_3 - q_4\,\frac{q_3-q_1}{q_4-q_1}\,(1-\dn_2^2)}
            {1-\frac{q_3-q_1}{q_4-q_1}\,(1-\dn_2^2)},
\end{array}
\]
where $\mathrm{ef}_i$ is a shorthand notation for appropriate
Jacobian function
(elliptic sine $\sn$, cosine $\cn$ or delta amplitude $\dn$)
\[
\mathrm{ef}_i = \mathrm{ef}\left[
    \sqrt{(q_4-q_2)(q_3-q_1)}\,\frac{t - t^*_i}{2},\kappa
\right]
\]
with modulus
\[
\kappa = \sqrt{\frac{(q_3-q_2)(q_4-q_1)}{(q_4-q_2)(q_3-q_1)}},
\]
and, two possible definitions of initial phases such that $q(0)=q_0$:
\[
\begin{array}{l}
\ds t^*_1 = \frac{-2}{\sqrt{(q_4-q_2)(q_3-q_1)}}\,
    \mathrm{F}\left[
    \arcsin\sqrt{\frac{(q_3-q_1)(q_0-q_2)}{(q_3-q_2)(q_0-q_1)}},\kappa
    \right],                                 \\[0.5cm]
\ds t^*_2 = \frac{2}{\sqrt{(q_4-q_2)(q_3-q_1)}}\,
    \mathrm{F}\left[
    \arcsin\sqrt{\frac{(q_4-q_2)(q_3-q_0)}{(q_3-q_2)(q_4-q_0)}},\kappa
    \right].
\end{array}
\]

\subsubsection{Action-angle variables and Pioncar\'e rotation number}

Now, we will relate continuous parametrization of invariant curves
given by Hamiltonian with the mapping equations.
First, we find the period of finite motion as
\[
\mathrm{T} \equiv \oint \dd\,t =
    2\,\int_{q_2}^{q_3}\frac{\dd\,q}{\sqrt{\mathcal{P}(q)}} =
    \frac{4\,\mathrm{K}[\kappa]}{\sqrt{(q_4-q_2)(q_3-q_1)}}.
\]
Then we calculate the time of one-step
\[
\begin{array}{l}
\ds \mathrm{T'} \equiv \int_0^\mathrm{T'} \dd\,t =
    \int_{q}^{q'}\frac{\dd\,q}{\sqrt{\mathcal{P}(q)}} =
    \int_{q_{2,3}}^{q_{2,3}'}\frac{\dd\,q}{\sqrt{\mathcal{P}(q)}}   \\[0.45cm]
\ds \,\,\!\quad = \frac{2\,\mathrm{F}[
        \arcsin\sqrt{
        \frac{q_{3,4}-q_{1,2}}{q_3-q_2}
        \frac{q_{2,3}(8-q_1+5\,q_{2,3}-q_{3,2}-q_4)}{4\,q_{1,4} + q_{2,3}\,(4+3\,q_{1,4}+q_{2,3}-q_{3,2}-q_{4,1})}
        }
    ,\kappa]}{\sqrt{(q_4-q_2)(q_3-q_1)}},
\end{array}
\]
where we can choose either one of $q_{2,3}$ as an initial coordinate,
since the integral does not depend on initial phase on a given
invariant curve, and, with the help of map equations we used
\[
    q_{2,3}' = p(q_{2,3}) = f(q_{2,3})/2 = -q_{2,3}\,\frac{q_{2,3}-a}{q_{2,3}+1}
    = \frac{q_{2,3}(q_1-q_{2,3}+q_{3,2}+q_4-4)}{4\,(q_{2,3}+1)}.
\]

\begin{figure}[t!]
    \centering
    \includegraphics[width=\linewidth]{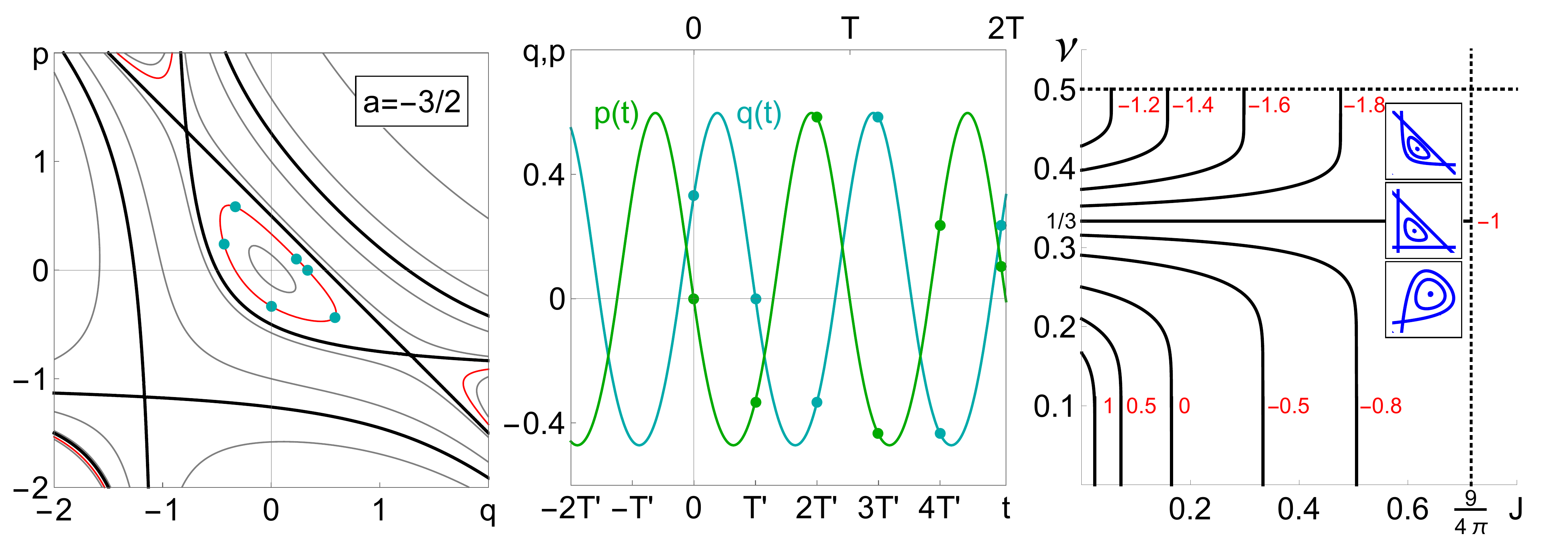}
    \caption{\label{fig:JNuSex}
    Illustration to parametrization of invariant curve and action-angle
    variables for the McMillan sextupole map.
    Left plot shows invariant level sets $\K_\mathrm{sex}(q,p)=\const$
    for $a=-3/2$ and $\Gamma=1$.
    Thick black curves are level sets containing separatrices, isolated
    fixed point and 2-cycle.
    Red level set is used to illustrate parametrization.
    Cyan points on the closed red curve are first five iterations of
    initial condition $(q_0,p_0)=(1/3,0)$.
    Middle plot shows continuous solution $q,p(t)$ of Hamiltonian system
    with $(q(0),p(0))=(q_0,p_0)$ and sampled at the rate $T'$ (time of
    one step);
    the time scale at the top is measured in periods of continuous system, $T$.
    Note that sampled points $(q_n,p_n)$ match corresponding points on
    the red level set from the left plot.
    Plot on the right is the action-angle variables for different values
    of $a$ (red labels).
    }
\end{figure}

Now we can rewrite the map in action-angle form
\[
\begin{array}{l}
    J' = J,                                 \\[0.25cm]
    \theta' = \theta + 2\,\pi\,\nu.
\end{array}
\]
The rotation number is defined as a ratio of the time of one-step
over the period {\bf[N]}
\[
    \nu = \mathrm{T}'/\mathrm{T} =
    \mathrm{F}\left[
        \arcsin\sqrt{
        \frac{q_3-q_1}{q_3-q_2}
        \frac{q_2(3\,q_2+2-a)}{2\,q_1(q_2+1) + q_2\,(q_2-a)}
        }
    ,\kappa\right]
    \Bigg/(2\,\mathrm{K}[\kappa]).
\]
The action is given by the integral 
\[
\begin{array}{l}
\ds J = \frac{1}{2\,\pi}\,\oint p\,\dd q =
    \frac{1}{2\pi}\,\int_{q_2}^{q_3} \frac{\sqrt{\mathcal{P}(q)}}{q+1}\,\dd q =    \\[0.35cm]
\ds \,\,\frac{q_2-q_1}{\pi}\,\sqrt{\frac{q_3-q_1}{q_4-q_2}}\,\Bigg\{
        \frac{a+4+q_4-q_1}{4}\,\mathrm{K}[\kappa] +
        \frac{a+4}{4}\frac{q_4-q_2}{q_2-q_1}\,\mathrm{E}[\kappa] -
        \\[0.35cm]
\ds \,\,\quad
        -\frac{a+1}{q_3-q_1}
            \,\Pi\left[\frac{q_4-q_2}{q_4-q_1}\,\kappa^2,\kappa\right]
        -\frac{(1+q_3)(1+q_4)}{q_3-q_1}
            \,\Pi\left[\frac{1+q_1}{1+q_2}\frac{q_4-q_2}{q_4-q_1}\,\kappa^2,\kappa\right]
    \Bigg\}.
\end{array}
\]
The initial phase can be defined as $\{\theta_0\} = \arctan(\{p_0\}/\{q_0\})$.
Right plot in Fig.~\ref{fig:JNuSex} shows rotation number as a
function of action for different values of $a$.
In addition, Fig.~\ref{fig:JNuSex} contains plots illustrating
parametrization of invariant curve.

\newpage

\subsection{Octupole map}

\subsubsection{Parametrization of individual curve}

For the octupole map, the parametrization of individual curves was
done by Iatrou and Roberts in~\cite{IR2002II}.
They also provided two different methods showing how an
arbitrary constant level set of an asymmetric McMillan map can be
transformed to canonical form, $\K_\mathrm{oct}(p,q)$.
Below we present their results using our notations and in the next
section we find canonical action-angle variables.

Due to additional symmetries of the invariant, the functional
dependence is much simpler compared to the sextupole case and all
trajectories can be parametrized in a form
\[
\left\{
\begin{array}{l}
    q_n = \sqrt[4]{\K}\,h(\kappa)\,\mathrm{ef}(u_0 + \eta_\pm\,n,\kappa),\\
    p_n = q_{n+1}
\end{array}
\right.
\]
where $\mathrm{ef}$ is an appropriate elliptic function with elliptic
modulus $\kappa$, $h(\kappa)$ is amplitude, $\eta_\pm$ is phase advance
for $a \gtrless 0$, and $u_0$ is initial phase defined by $q_0$.

For $\Gamma < 0$, appropriate elliptic function for stable
trajectories is elliptic sine, $\sn$.
When $\Gamma > 0$ one should use elliptic cosine, $\cn$, or delta
amplitude, $\dn$, for trajectories rounding the origin or
trajectories inside the lemniscate respectively.
Expressions for all parameters are listed in Table~\ref{tab:IR}
below (with $\eta_- = 2\mathrm{K}[\kappa] - \eta_+$).

\subsubsection{Action-angle variables and Pioncar\'e rotation number}

Note that in~\cite{IR2002II}, the authors also defined action-angle
variables which are not canonical:
elliptic modulus $\kappa$ is used as an action instead of
an area under the curve and phase $\eta$ is used instead of
a rotation number.

The rotation number for trajectories rounding the origin or
lemniscate (i.e. all $\sn$ and $\cn$ cases)
\[
    \nu = \frac{\eta}{4\,\mathrm{K}[\kappa]}
\]
and for trajectories inside lemniscate orbiting one of the symmetric
fixed points $\zeta^{(1-2,3)}$ ($\dn$ case)
\[
    \nu = \frac{\eta}{2\,\mathrm{K}[\kappa]}.
\]

\begin{table*}[t!]
\begin{tabular}{p{3cm}p{4.5cm}p{4.5cm}p{4.5cm}}
\hline
\hline
    & cn    & dn    & sn			                                \\\hline
								                                    \\[-0.3cm]
$\K\in$                                                             &
$[0;\infty) \geq 0$							                        &
$\left[-(|\frac{a}{2}|-1)^2;0\right] \leq 0$				        & 
$\left[0; (|\frac{a}{2}|-1)^2\right] \geq 0$			            \\[0.3cm]
$B(a,\K)$									                        &
$\displaystyle \frac{(a/2)^2-1+\K}{\sqrt{|\K|}}$                    &
$\displaystyle \frac{(a/2)^2-1-\K}{\sqrt{|\K|}}$                    &
$\displaystyle-\frac{(a/2)^2-1-\K}{\sqrt{|\K|}}$                    \\[0.4cm]
                                                                    &
$\displaystyle =\frac{k}{k'}-\frac{k'}{k}\in \mathbb{R}$            &
$\displaystyle =\frac{1}{k'}+k'\geq 2$					            &
$\displaystyle =\frac{1}{k} +k \geq 2$				                \\[0.4cm]
$k(B)$									                            &
$\displaystyle \frac{1}{\sqrt{2}}\sqrt{1+\frac{B}{\sqrt{B^2+4}}}$	&
$\displaystyle \frac{\sqrt{B+2}-\sqrt{B-2}}{2\,(B^2-4)^{-1/4}}$	    &
$\displaystyle \frac{B-\sqrt{B^2-4}}{2}$			                \\[0.4cm]
$k'(B)$									                            &
$\displaystyle \frac{1}{\sqrt{2}}\sqrt{1-\frac{B}{\sqrt{B^2+4}}}$	&
$\displaystyle \frac{B-\sqrt{B^2-4}}{2}$			                &
$\displaystyle \frac{\sqrt{B+2}-\sqrt{B-2}}{2\,(B^2-4)^{-1/4}}$	    \\[0.4cm]
$\eta_+$									                        &
$\displaystyle \ads\,\frac{\sqrt{k\,k'}}{\sqrt[4]{|\K|}}$		    &
$\displaystyle \acs\,\frac{\sqrt{k'   }}{\sqrt[4]{|\K|}}$		    &
$\displaystyle \ans\,\frac{\sqrt{k    }}{\sqrt[4]{|\K|}}$	        \\[0.4cm]
$h(\kappa)$							                                &
$\displaystyle \sqrt{k/k'}$		                                    &
$\displaystyle \sqrt{1/k'}$		                                    &
$\displaystyle \sqrt{k}$	                                        \\[0.3cm]
$u_0$									                            &
$\displaystyle \acn\,\frac{q_0}{h(\kappa)\,\sqrt[4]{|\K|}}$         &
$\displaystyle \adn\,\frac{q_0}{h(\kappa)\,\sqrt[4]{|\K|}}$         &
$\displaystyle \asn\,\frac{q_0}{h(\kappa)\,\sqrt[4]{|\K|}}$         \\[0.4cm]
\hline
\hline
\end{tabular}
\caption{Elliptic parametrization of
        stable trajectories of McMillan
        octupole map (after~\cite{IR2002II});
        $\eta_- = 2\mathrm{K}[\kappa] - \eta_+$.}
\label{tab:IR}
\vspace{-0.5cm}
\end{table*}

Using parametrization of individual curves, we now can compute the
action integral~\cite{zolkin2016analytical} as
\[
\ds J   = \frac{1}{2\,\pi}\,\oint p\,\dd q
        = \frac{\sqrt{|\K|}}{2\,\pi} \times\left\{
        \begin{array}{l}
        \ds    \qquad\kappa  \,\,S_\sn,   \\[0.2cm]
        \ds    (\kappa/\kappa') \,S_\cn,     \\[0.2cm]
        \ds    (1/\kappa')      \,S_\dn,
        \end{array}
        \right.
\]
where $S_\mathrm{ef}$ are areas of {\it elliptic Lissajous
curves}
with equal frequencies and phase difference $\eta$
(see Appendix~\ref{sec:AppLiss}):
\begin{eqnarray*}
J_{\text{sn}} & = &
\frac{2}{\pi}\frac{\sqrt{|\K|}}{\kappa}\frac{1}{\sn^3 \eta}\Big\{
	- \dn^2 \eta                \,\mathrm{K}[\kappa]
	+ \sn^2 \eta                \,\mathrm{E}[\kappa]
	+ \cn^2 \eta\,\dn^2 \eta    \,\Pi[k^2\sn^2 \eta,\kappa]
\Big\},
\\
J_{\text{cn}} & = &
\frac{2}{\pi}\frac{\sqrt{|\K|}}{\kappa\,\kappa'}\frac{\dn\,\eta}{\sn^3 \eta}\Big\{
	                              \mathrm{K}[\kappa]
	- \sn^2 \eta                \,\mathrm{E}[\kappa]
	- \cn^2 \eta                \,\Pi[k^2\sn^2 \eta,\kappa]
\Big\},
\\
J_{\text{dn}} & = &
\frac{1}{\pi}\frac{\sqrt{|\K|}}{\kappa'}\frac{\cn\,\eta}{\sn^3 \eta}\Big\{
	  (\sn^2 \eta+\dn^2 \eta)   \,\mathrm{K}[\kappa]
	- \sn^2 \eta                \,\mathrm{E}[\kappa]
	- \dn^2 \eta                \,\Pi[k^2\sn^2 \eta,\kappa]
\Big\}.
\end{eqnarray*}
Figure \ref{fig:JNuOct} below shows action-angle
variables for McMillan octupole map for different
values of parameters $a$ and $\Gamma$.

\begin{figure}[h!]
    \centering
    \includegraphics[width=0.78\linewidth]{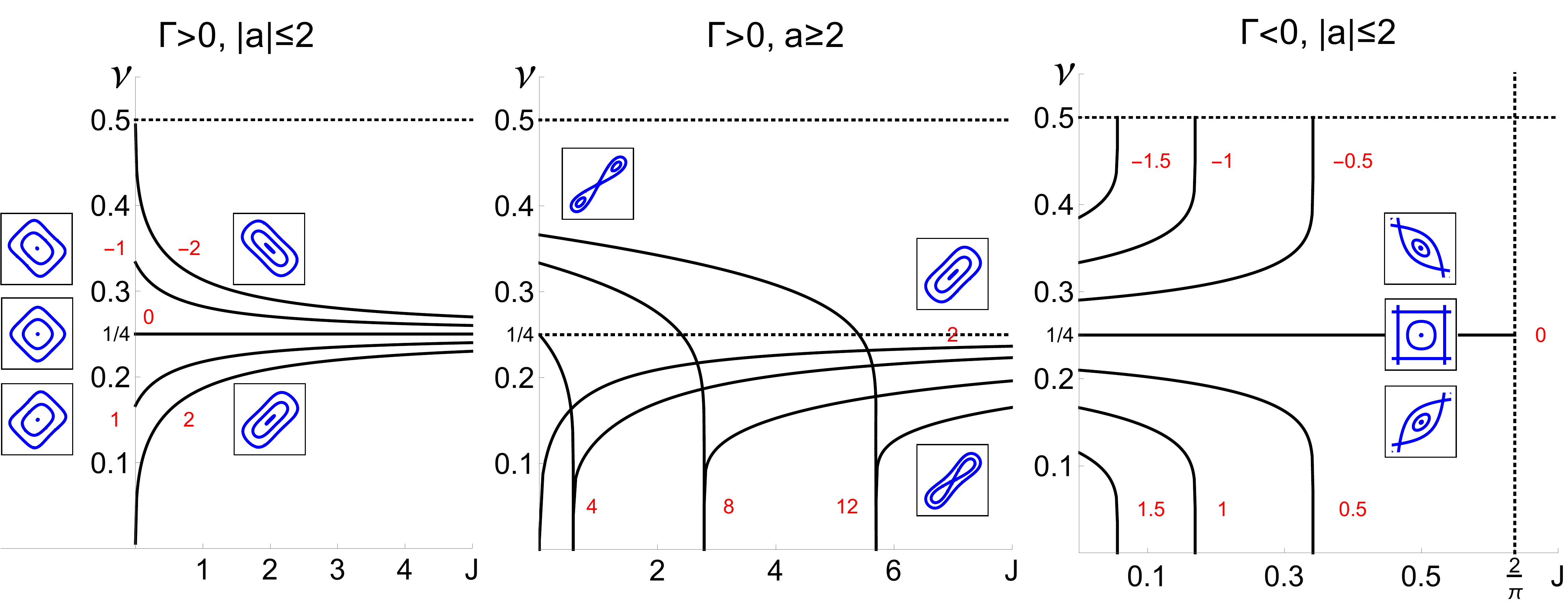}
    \caption{\label{fig:JNuOct}
    Action-angle variables for the McMillan octupole map for
    different values of $a$ (red labels) and $|\Gamma|=1$.
    Each plot corresponds to a different regime of oscillations:
    the left plot shows regime with single fixed point,
    the one in the middle is regime with lemniscate,
    and regime with finite dynamical aperture ($\Gamma < 0$)
    is on the right.
    }
\end{figure}

\section{\label{sec:Remarks}Applications of McMillan map}

Besides the obvious importance of McMillan map as one of very few
known exact integrable models (e.g. see~\cite{zolkin2022machine}
for the short review on integrability of symplectic maps of the
plane), it has a much deeper meaning.
In order to see that, we first should comment on a McMillan-Turaev
form of the map
\[
\begin{array}{ll}
    \mathcal{M}_\mathrm{MT}:    & q' = p,       \\[0.25cm]
                                & p' =-q + f(p).
\end{array}
\]
At first glance the map might look very restrictive;
McMillan's original idea was based on its simplicity and clear
symmetries, while still having some degree of freedom in modifying
the dynamics by varying only $f(p)$.
As it was demonstrated later by D. Turaev~\cite{turaev2002polynomial},
any symplectic $2n$-D mapping can be approximated to the MT form
up to a given numerical precision.
Thus, in a certain sense, MT form is the canonical form for all
symplectic maps with differentiable $f(p)$.
In next few subsections we show that McMillan sextupole and
octupole maps are at the core of general MT map and will propose
a natural extension of the optical function formalism used in
accelerator physics.
While material below requires its own publication, which we will
make soon, we briefly describe main ideas.

\subsection{\label{sec:PertTh}McMillan map and perturbation theory}

Consider the general map in MT form with $f(p)$ being differentiable
function of its argument.
Without loss of generality we will assume the fixed point to be at
the origin, which requires $f(0)=0$.
For an arbitrary otherwise $f(p)$ the global dynamics is known to
be chaotic, except very special integrable cases
~\cite{mcmillan1971problem,suris1989integrable,zolkin2022machine}.
As a consequence of KAM theorem~\cite{kolmogorov1954conservation,moser1962invariant,arnol1963small},
if linear part of the map is stable
\[
	\left| \pd_p f(0) \right| \leq 2,
\]
then there exist an area of quasi-integrable motion around fixed
point.
Now, we shall introduce a small positive parameter $\varepsilon$
characterizing the amplitude of oscillations.
It can be done using a change of variables
$(q,p) \rightarrow \varepsilon\,(q,p)$:
\[
\begin{array}{l}
	q' = p							\\[0.25cm]
\ds	p' =-q + \frac{1}{\varepsilon}\,f(\varepsilon\,p) =
	    -q + a\,p	+ \varepsilon\,\frac{b}{2!}\,p^2
			+ \varepsilon^2\,\frac{c}{3!}\,p^3
			+ \ldots.
\end{array}
\]
where we expanded the force function in a power series of
$(\varepsilon\,p)$ and
\[
	a \equiv \pd_p   f(0),\qquad
	b \equiv \pd_p^2 f(0),\qquad
	c \equiv \pd_p^3 f(0),\qquad
	\ldots.
\]

After the introduction of $\varepsilon$, one can look for an
approximated invariant of motion expanded in powers of a small
parameter.
In this case, we require the invariant to be conserved to an
accuracy of  order $O(\varepsilon^{n+1})$:
\begin{equation}
\label{math:KK'}
	\K^{(n)}(p',q') - \K^{(n)}(p,q) = O(\varepsilon^{n+1}).
\end{equation}
The invariant can be sought in the form of a polynomial
\[
\K^{(n)} = \K_0 + \varepsilon\,\K_1 + \varepsilon^2\,\K_2 + \ldots +
\varepsilon^n\,\K_n,
\]
with $\K_m$ being homogeneous polynomials in $p$ and $q$ of
$(m+2)$ degree:
\[
\begin{array}{l}
	\K_0 =  C_{2,0}\,p^2    +
	        C_{1,1}\,p\,q\, +
	        C_{0,2}\,q^2,			    \\[0.25cm]
	\K_1 =  C_{3,0}\,p^3    +
	        C_{2,1}\,p^2 q  +
	        C_{1,2}\,p\,q^2 +
	        C_{0,3}\,q^3,               \\[0.25cm]
	\cdots,
\end{array}
\]
and $C_{i,j}$ are coefficients to be determined in order to
satisfy Eq.~(\ref{math:KK'}).
Leaving it to a reader to verify, we will provide a general result.
In first order of such a perturbation theory we have
\begin{equation}
\label{math:KsexAp}    
    \K(p,q) =   p^2 - a\,p\,q + q^2 -
                \frac{b}{2!}\,\frac{p\,q\,(p+q)}{a-1}\,\varepsilon +
                O(\varepsilon^2).
\end{equation}
In the case if $f(p)$ is an even function of its argument (or just
$b=0$), the expansion begins from the next order and we have
\begin{equation}
\label{math:KoctAp}
    \K(p,q) =   p^2 - a\,p\,q + q^2 -
                \frac{c}{3!}\,\frac{p^2q^2}{a}\,\varepsilon^2 +
                C\,(p^2 - a\,p\,q + q^2)^2\varepsilon^2 +
                O(\varepsilon^3),
\end{equation}
where $C$ is a coefficient such that Eq.~(\ref{math:KK'}) is
satisfied for any value of $C$.

Here one can recognize (if we set $C=0$ in Eq.~(\ref{math:KoctAp}))
that these are invariants of McMillan sextupole and octupole maps.
Thus we learn that these McMillan mappings, are not just two
integrable systems;
in the same way the quadratic invariant
\[
    \K^{(0)}(p,q) = p^2 - a\,p\,q + q^2
\]
and linear force $f(p) = a\,p$ describe the dynamics around the
fixed point of a general dynamical systems in the zeroth order
(known as linearization), McMillan sextupole and octupole maps
with invariants $\K_\mathrm{sex,oct}$ provide the next order
approximation.
As a result one can ``integrate out'' additional dynamical
properties, such as action-angle dependence and even boundary
of stability around some major resonances.
After the truncation of force function at the second order one
produces new dynamical system which is chaotic again (quadratic
or cubic H\'enon maps).
However addition of higher order terms according to
Eqs.~(\ref{math:fsexExp}) and (\ref{math:foctExp}) matches
the force with invariant producing nonlinear integrable map.

Another famous chaotic dynamical system we would like to use as an
example is
{\it Chirikov map}~\cite{chirikov1969research,chirikov1979universal}
(also known as {\it Chirikov standard} or {\it Chirikov-Taylor map}).
In accelerator physics it is used to model longitudinal dynamics
in a ring with single thin RF station, and in MT form is given by
\[
\begin{array}{l}
    q' = p,       \\[0.25cm]
    p' =-q + 2\,p - K\,\sin p.
\end{array}
\]
This map has an approximated invariant in the octupole
form~(\ref{math:KoctAp}) with $a=2-K$ and $c = K$.

\subsection{\label{sec:Twiss}Nonlinear Twiss parameters}


Now let us consider a simple accelerator lattice with 1 degree
of freedom consisting of elements of linear optics (drift spaces,
dipoles and quadrupoles) and a single thin nonlinear lens:
\[
      \mathrm{F}:\quad\,\,
      \begin{bmatrix}
      x \\ \dot{x}
      \end{bmatrix}' =
      \begin{bmatrix}
      x \\ \dot{x}
      \end{bmatrix} +
      \begin{bmatrix}
      0 \\ F(x)
      \end{bmatrix}.
\]
The action on a test particle from all linear elements can be
represented using matrix with Courant-Snyder
parametrization~\cite{courant1958theory}
\[
      \mathrm{M}:\quad
      \begin{bmatrix}
      x \\ \dot{x}
      \end{bmatrix}' =
      \begin{bmatrix}
      \cos \Phi + \alpha\,\sin \Phi	& \beta\,\sin \Phi		\\
      -\gamma\,\sin\Phi		& \cos \Phi - \alpha\,\sin \Phi
      \end{bmatrix}
      \begin{bmatrix}
      x \\ \dot{x}
      \end{bmatrix},
\]
where $\alpha$, $\beta$ and $\gamma$ are {\it Twiss parameters}
(also known as {\it Courant-Snyder parameters}) at the thin
lens location, and, $\Phi$ is the {\it betatron phase advance}
over linear optics insert
\[
    \Phi = \int \frac{\dd s}{\beta(s)}.
\]
Without nonlinear lens Twiss parameters are functions of
longitudinal coordinate $s$ with $\beta(s)$ refereed as
$\beta$-function, $\alpha(s) \equiv -\dot{\beta}(s)/2$ and
$\gamma(s) \equiv [1+\alpha^2(s)]/\beta(s)$.
At any location the {\it Courant-Snyder invariant} is defined as
\[
    \gamma(s)\,x^2(s) + 2\,\alpha(s)\,x(s)\,\dot{x}(s) + \beta(s)\,\dot{x}^2(s) =
    \const
\]
and rotation number (or {\it betatron tune} in accelerator
physics) is independent of amplitude and given by
\[
    \nu = \frac{1}{2\,\pi}\,\oint\frac{\dd s}{\beta(s)}.
\]

When the nonlinear lens is introduced, the combined one-turn map
$\mathrm{M}\circ\mathrm{F}$ can be rewritten in MT form using
change of variables
\[
\left\{
\begin{array}{l}
	q = x,		\\[0.2cm]
	p = x\,(\cos \Phi + \alpha\,\sin \Phi) + \dot{x}\,\beta\,\sin \Phi,
\end{array}\right.
\]
with force function being
\[
f(q) = 2\,q\,\cos\Phi + \beta\,\sin\Phi\,F(q).
\]
Using results from the previous subsection we know that we can
define an approximated invariant
\[
    \K(p,q) \approx \mathrm{C.S.} -
                    \frac{b}{2!}\,\frac{\Pi\,\Sigma}{a-1}
\]
or if $b = 0$
\[
    \K(p,q) \approx \mathrm{C.S.} -                 
                    \frac{c}{3!}\,\frac{\Pi^2}{a},
\]
where we used our symmetric notations
$\mathrm{C.S.} \equiv p^2 - a\,p\,q + q^2$,
$\Pi \equiv p\,q$ and $\Sigma \equiv p+q$ with 
\[
    a = 2\,\cos\Phi + \beta\,\sin\Phi\,\pd_q\,F(0),\qquad
    b = \beta\,\sin\Phi\,\pd_{qq }\,F(0),\qquad
    c = \beta\,\sin\Phi\,\pd_{qqq}\,F(0).
\]
New invariant can be easily propagated through the linear part
of lattice and defined for any azimuth, $\K(p,q;s)$, and if
needed $(q,p)$ can be inverted to physical variables
$(x,\dot{x})$.
Now one can see that coefficients in polynomial invariant can be
seen as nonlinear analog of Twiss parameters now including higer
order terms.
While physical variables are convenient for the representation of
final results, all calculations are more convenient in MT form
due to the fact that it has clear symmetries, small amount of
terms and the only parameters used are intrinsic parameters,
such as rotation number at the origin (or $a$) or location of
singularities $\Gamma$;
MT form and corresponding symmetric invariant is in a way similar
to use of Floquet variables, when we remove redundancy in optical
functions (since $\alpha$ and $\gamma$ are defined through $\beta$-function) and use physical variables after we propagate
the system in normalized coordinates.
We also should briefly note that one can incorporate effect from
both, sextupole and octupole, components of nonlinear lens
\[
    \K(p,q) \approx \mathrm{C.S.} -
                    \frac{b}{2!}\,\frac{\Pi\,\Sigma}{a-1} +
                    \left[
                    \left(\frac{b}{2!}\right)^2 - \frac{c}{3!}
                    \right]
                    \,\frac{\Pi^2}{a\,(a-1)},
\]
resulting in an additional parameter --- the ratio of nonlinear
components $b/c$.
In this case we have a more general symmetric McMillan map
invariant level sets of which can be transformed to octupole
McMillan map (see~\cite{IR2002II}).

While conventional Courant-Snyder formalism been successfully used
by accelerator community for years, we should remember that it is
only linearization ignoring any nonlinear effects and definitely
have room for improvement.
Suggested nonlinear extension based on McMillan integrable maps has
following advantages:
\begin{itemize}
    \item In traditional Courant-Snyder formalism the rotation
    number (betatron tune) is independent of amplitude, while
    in its nonlinear extension we have $\nu(J) \neq \const$;
    it helps to incorporate chromatic effect of sextupoles and
    tune shift due to octupoles.
    \item While linear Courant-Snyder invariant is stable for
    any amplitude, McMillan invariant introduces dynamical aperture
    which is quite accurate around resonances of 1-st, 2-nd, 3-rd
    and 4-th order, respectively $a=2,-2,-1$ and 0.
    In particular the last two resonances plays critical role in
    sextupole and octupole resonant injection/extraction and
    new theory provides quick and accurate location of unstable
    points and separatrices of motion.
    A specific example is given in next subsection.
    \item The new formalism is very natural:
    original invariant is in polynomial form and non-linearities
    are consistently incorporated.
    In addition, the new extended invariant again corresponds to
    exactly integrable system and familiar to accelerator
    community.
\end{itemize}

\subsection{\label{sec:AccExample}General accelerator lattice}

In previous subsection we used a model accelerator with only one
nonlinear lens.
As a result, we had an ability to analytically convert a one turn
map to MT form.
But what should we do in the case of general symplectic map of the
plane representing more realistic accelerator with multiple nonlinear
thin lenses located at different positions along the lattice?
Note that thick lenses can always be incorporated into analysis.
If for each element of the lattice an exact transformation is known,
then one simply expand the full one-turn map into a power series
up to the third order, and, if some transformations are unknown
one can use symplectic integrator~\cite{yoshida1990construction}
consisting of drifts and thin kicks or Dirac interaction picture 
along with the Magnus expansion~\cite{morozov2017dynamical}.

The first method is to numerically transform the map to MT form
using Turaev theorem and thus reduce the analysis to symmetric
McMillan mappings.
However, one can use the perturbation theory described above for
any symplectic map;
the only difference is that instead of symmetric one should use
general polynomials for the approximated invariant of motion.
This approach is less practical in a scenes that asymmetric
polynomials have too many coefficients/nonlinear optical functions
(4 for sextupole and 5 for octupole).
Also note that in general case after the transformation to MT form,
the contribution from all sextupoles (and/or sextupole components
from other nonlinear lenses) combined in one term of force function;
same is true for the octupole component.
As reader can guess, in zeroth order one obtain Courant-Snyder
invariant, and in first asymmetric McMillan map.
This again allows for a quick extraction of aforementioned
dynamical properties including structure of the phase space.

\begin{figure}[h!]
    \centering
    \includegraphics[width=\linewidth]{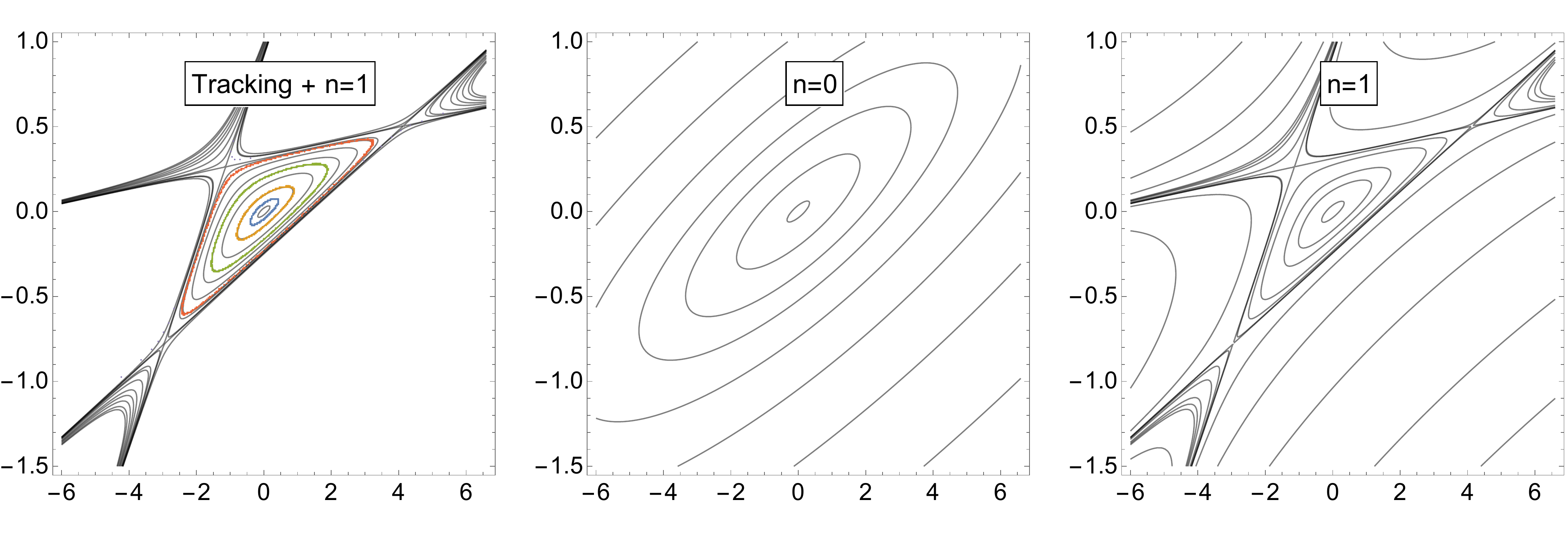}
    \caption{\label{fig:Mu2e}
    Left plot provides tracking of the third-integer resonant extraction
    in horizontal plane for the Mu2e experiment at Fermilab (colored
    dots);
    black level sets correspond to nonlinear Courant-Snyder invariant
    of the first order (shown separately in right plot).
    Middle and right plots show constant level sets of linear and
    nonlinear Courant-Snyder invariants, $\K^{(0)}(x,\dot{x})$ and
    $\K^{(1)}(x,\dot{x})$.
    All plots are made in physical coordinates $(x,\dot{x})$.
    }
\end{figure}

In order to illustrate the second approach described above, we will
use the example of a third-integer resonant extraction for the Mu2e
experiment at Fermilab.
Since beam is flat, we will apply our one dimensional theory to the
dynamics in horizontal plane $(x,\dot{x})$.
After producing a one-turn map including all sextupoles and linear
optics elements, we expand it in power series and compute zeroth and
first order approximated invariants:
\[
\begin{array}{l}
\K^{(0)}(x,\dot{x}) =
  7.29976   \,\dot{x}^2
- 2.16717   \,\dot{x} x
+ 0.250363  x^2,            \\[0.25cm]
\K^{(1)}(x,\dot{x}) =
\K^{(0)}(x,\dot{x}) 
- 7.24291   \,\dot{x}^3 
+ 5.85441   \,\dot{x}^2 x
- 1.00907   \,\dot{x} x^2
+ 0.0338591 \,x^3.
\end{array}
\]
$K^{(0)}(x,\dot{x})$ is the Courant-Snyder invariant and
$K^{(1)}(x,\dot{x})$ is in the form of asymmetric McMillan
sextupole map.
Fig.~\ref{fig:Mu2e} provides comparison between tracking (using
full one-turn map) and approximated invariants.
As one can see the dynamic aperture (location of unstable fixed
points) and other main features (such as separatrices and shape
of stable phase space trajectories) match quite well;
without going into details here we will mention that action-angle
dependency is also within good accuracy.
In an upcoming soon article we will consider in details the
application of this theory specifically to accelerator physics
considering the range of applicability and several examples
demonstrating the validity of estimated dynamical properties.

\section{Summary}

In this article, we revisited an integrable McMillan map and showed
that it is at the core of general symplectic dynamics of the plane.
With the help of perturbation theory, we demonstrated that McMillan
sextupole and octupole mappings are the first and second order
approximations for planar transformations in McMillan-Turaev
form.
In particular, this allows the natural extension of the optical
function formalism used in accelerator physics;
this new formalism incorporates the chromatic effect of sextupoles
and tune shift due to octupole magnets, along with dynamical
aperture around low order resonances.
As an example, we used a real accelerator lattice used for the
Mu2e experiment at Fermilab, which revealed a great correspondence
between tracking and perturbation theory applied for the
third-order resonant extraction.

This manuscript is first in a series of publications dedicated to
McMillan map and related integrable systems in higher dimensions
with its applications to accelerator physics.
In this work, we provide a systematic description of basic
underlying ideas: in particular, we give a complete description of
all stable trajectories, including bifurcation diagrams,
parametrization of invariant curves (see~\cite{IR2002II}),
Pioncar\'e rotation numbers, and canonical action-angle variables.

\section{Acknowledgments}

The authors would like to thank Eric Stern (FNAL) and Taylor Nchako
(Northwestern University) for carefully reading this manuscript and
for their helpful comments.
Moreover, we would like to extend our gratitude to Vladimir Nagaslaev
(FNAL) for multiple discussions and his generous contributions in
preparation of Fig.~\ref{fig:Mu2e}.

This research is supported by Fermi Research Alliance, LLC under
Contract No. DE-AC02-07CH11359 with the U.S. Department of Energy
and by the University of Chicago.

\appendix

\newpage
\section{\label{sec:AppStab}Stability of fixed points and 2-cycles
        for \\ McMillan sextupole and octupole mappings}

In this appendix we provide tables with stability domain for fixed
points and two-cycles of the integrable McMillan sextupole and
octupole mappings.

\begin{table}[h!]
\begin{tabular}{l|ccc}
                & $\K(\zeta)$
    & Domain $\zeta \in \mathbb{R}^2$
    & Stability domain                              \\\hline
                &             &        &            \\[-0.25cm]
$\zeta^{(1-1)}$ & 0
    & $\mathbb{R}^2$
    & $(\Gamma+\epsilon)(\Gamma-\epsilon) > 0$      \\[0.25cm]
&   & \includegraphics[width=3.8cm]{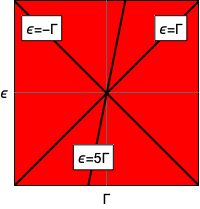}
    & \includegraphics[width=3.8cm]{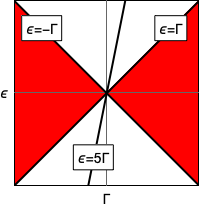}      \\[0.25cm]
$\zeta^{(1-2)}$ & $\frac{8}{27}\,(\Gamma+\epsilon)^3$
    & $\mathbb{R}^2$
    & $(\Gamma+\epsilon)(5\,\Gamma-\epsilon) < 0$   \\[0.25cm]
&   & \includegraphics[width=3.8cm]{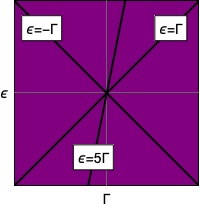}
    & \includegraphics[width=3.8cm]{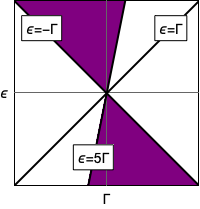}      \\[0.25cm]
$\zeta^{(2)}$   & $4\,\Gamma\,(\Gamma-\epsilon)^2$
    & $(\Gamma-\epsilon)(5\,\Gamma-\epsilon) > 0$
    & $(\Gamma-\epsilon)(5\,\Gamma-\epsilon) < 0$   \\[0.25cm]
&   & \includegraphics[width=3.8cm]{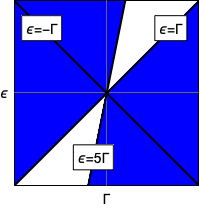}
    & \includegraphics[width=3.8cm]{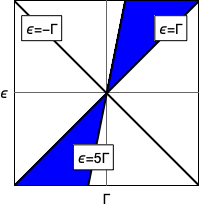}
\end{tabular}
\caption{\label{tab:SexStab}
    Value of the invariant, $\K_\text{sex}(\zeta)$, domain where point is real and its stability domain for fixed points and two-cycles of the McMillan
    sextupole map.
}
\end{table}

\begin{table}[p!]
\begin{tabular}{l|ccc}
                & $\K(\zeta)$
    & Domain $\zeta \in \mathbb{R}^2$
    & Stability domain                              \\\hline
                &             &        &            \\[-0.25cm]
$\zeta^{(1-1)}$ & 0
    & $\mathbb{R}^2$
    & $(\Gamma+\epsilon)(\Gamma-\epsilon) > 0$      \\[0.25cm]
&   & \includegraphics[width=3.8cm]{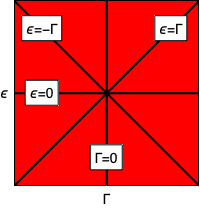}
    & \includegraphics[width=3.8cm]{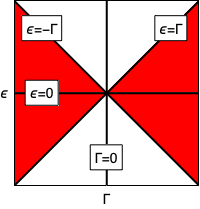}      \\[0.25cm]
$\zeta^{(1-2,3)}$ & $-(\Gamma+\epsilon)^2$
    & $\epsilon <-\Gamma$
    & $\Gamma\,(\Gamma+\epsilon) < 0$   \\[0.25cm]
&   & \includegraphics[width=3.8cm]{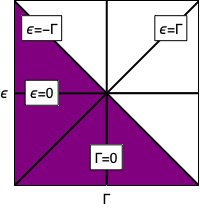}
    & \includegraphics[width=3.8cm]{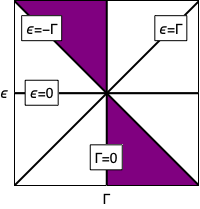}      \\[0.25cm]
$\zeta^{(2)}$   & $-(\Gamma-\epsilon)^2$
    & $\epsilon > \Gamma$
    & $\Gamma\,(\Gamma-\epsilon) < 0$   \\[0.25cm]
&   & \includegraphics[width=3.8cm]{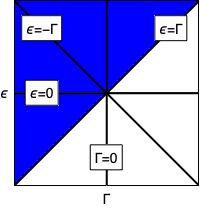}
    & \includegraphics[width=3.8cm]{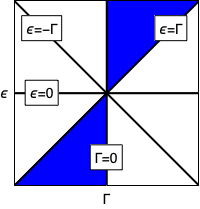}
\end{tabular}
\caption{\label{tab:OctStab}
    Value of the invariant, $\K_\text{oct}(\zeta)$, domain where point is real and its stability domain for fixed points and two-cycles of the McMillan
    octupole map.
}
\end{table}

\newpage
\section{\label{sec:AppLiss}Elliptic Lissajous figures}

Traditionally a {\it Lissajous figure} (or curve) is the graph
defined by the parametric set of equations
\[
\begin{array}{l}
    x = \sin(\omega_1\,t),         \\[0.2cm]
    y = \sin(\omega_2\,t+\eta).
\end{array}
\]
A natural extension is substitution of Jacobi elliptic functions
instead of trigonometric
\[
\begin{array}{l}
    x = \mathrm{ef}\,(\omega_1\,t,\kappa),         \\[0.2cm]
    y = \mathrm{ef}\,(\omega_2\,t+\eta,\kappa).
\end{array}
\]
We will define graph of this parametric set as
{\it Lissajous-Jacobi curve} or {\it elliptic Lissajous figure};
one can think of even further generalization by using two different
elliptic functions, e.g. $\sn$ vs. $\cn$, which we will not consider.

\begin{figure}[t!]
    \centering
    \includegraphics[width=\linewidth]{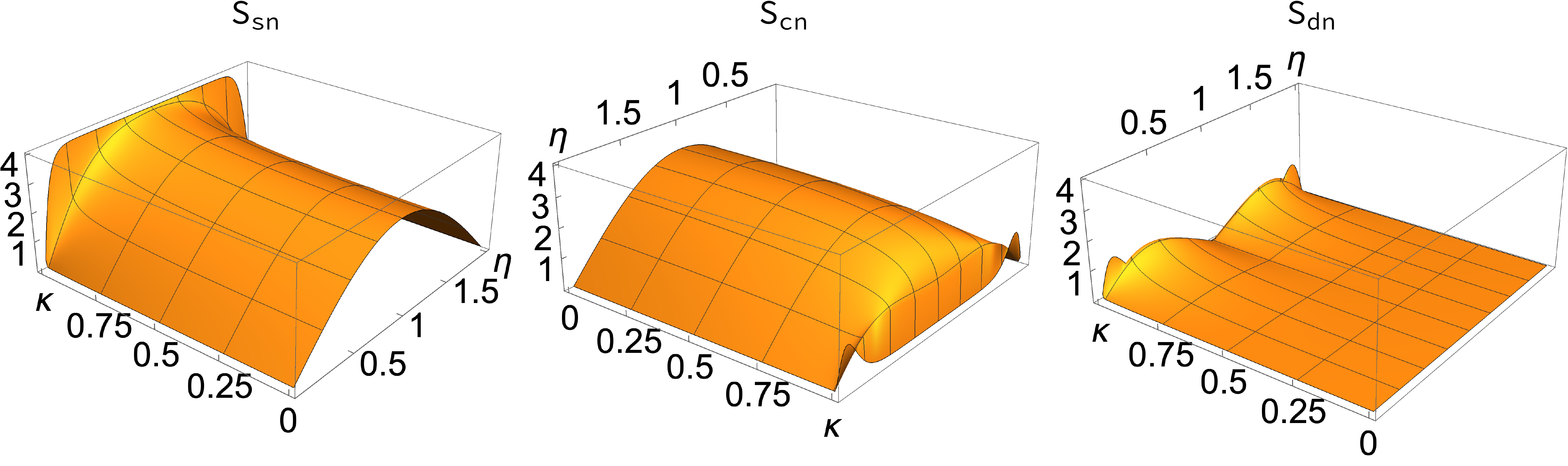}
    \caption{\label{fig:LssjArea}
    Area under the closed elliptic Lissajous curve for major elliptic
    functions and unit ratio ($\omega_1:\omega_2 = 1$) as a function
    of modulus $\kappa$ and phase difference $\eta$.
    Note that $\eta$ is measured in units of quarter period
    $\mathrm{K}[\kappa]$.
    }
\end{figure}

\begin{figure}[p]
    \centering
    \includegraphics[width=0.7\linewidth]{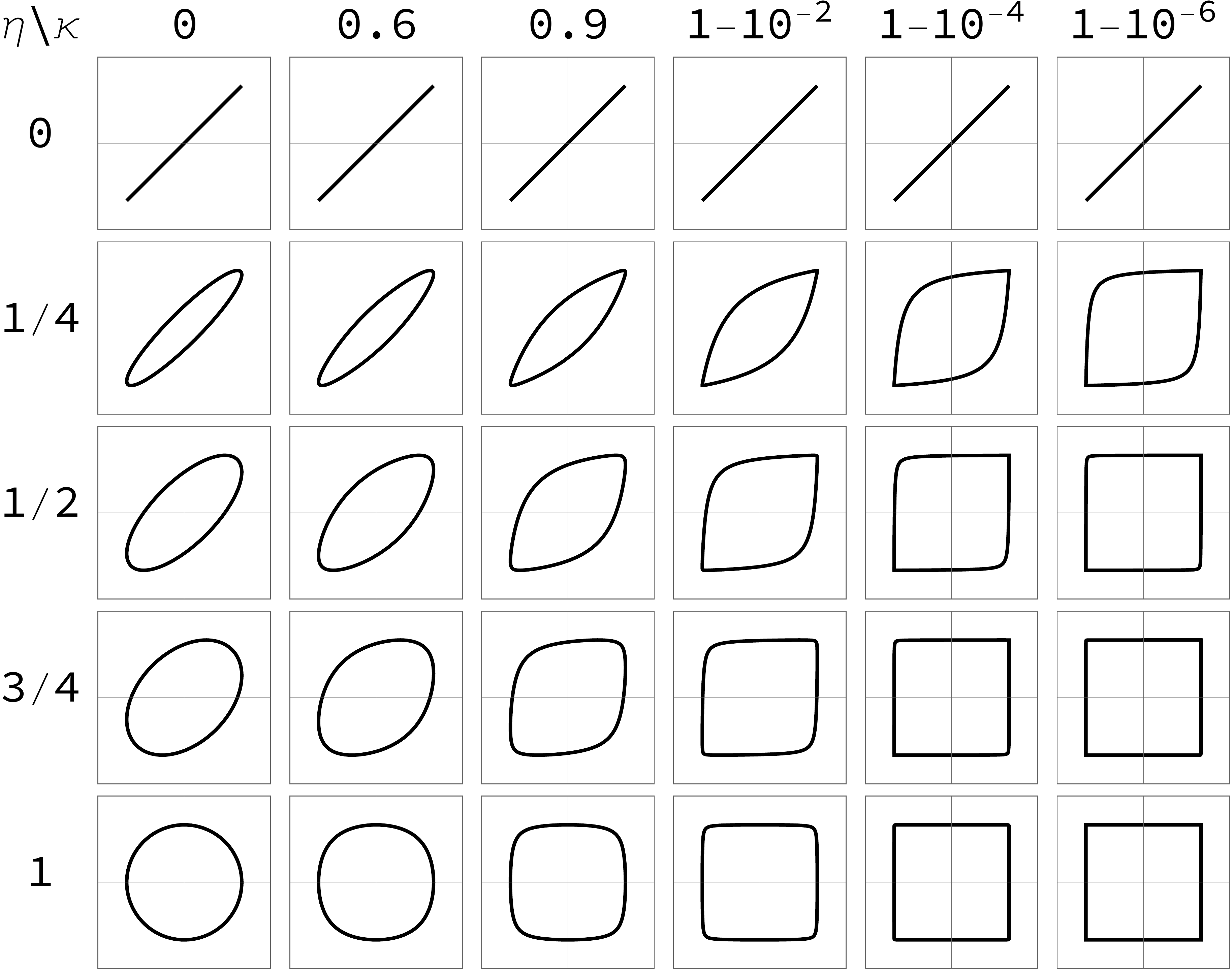}
    \caption{\label{fig:LssjSN}
    Lissajous-Jacobi figures ($\sn$ vs. $\sn$) for a unit ratio,
    $\omega_1:\omega_2 = 1$.
    Columns and rows corresponds to different values of elliptic
    modulus $\kappa$ and phase difference $\eta$.
    Note that $\eta$ is measured in units of quarter period
    $\mathrm{K}[\kappa]$.
    }
\end{figure}

\begin{figure}[p]
    \centering
    \includegraphics[width=0.7\linewidth]{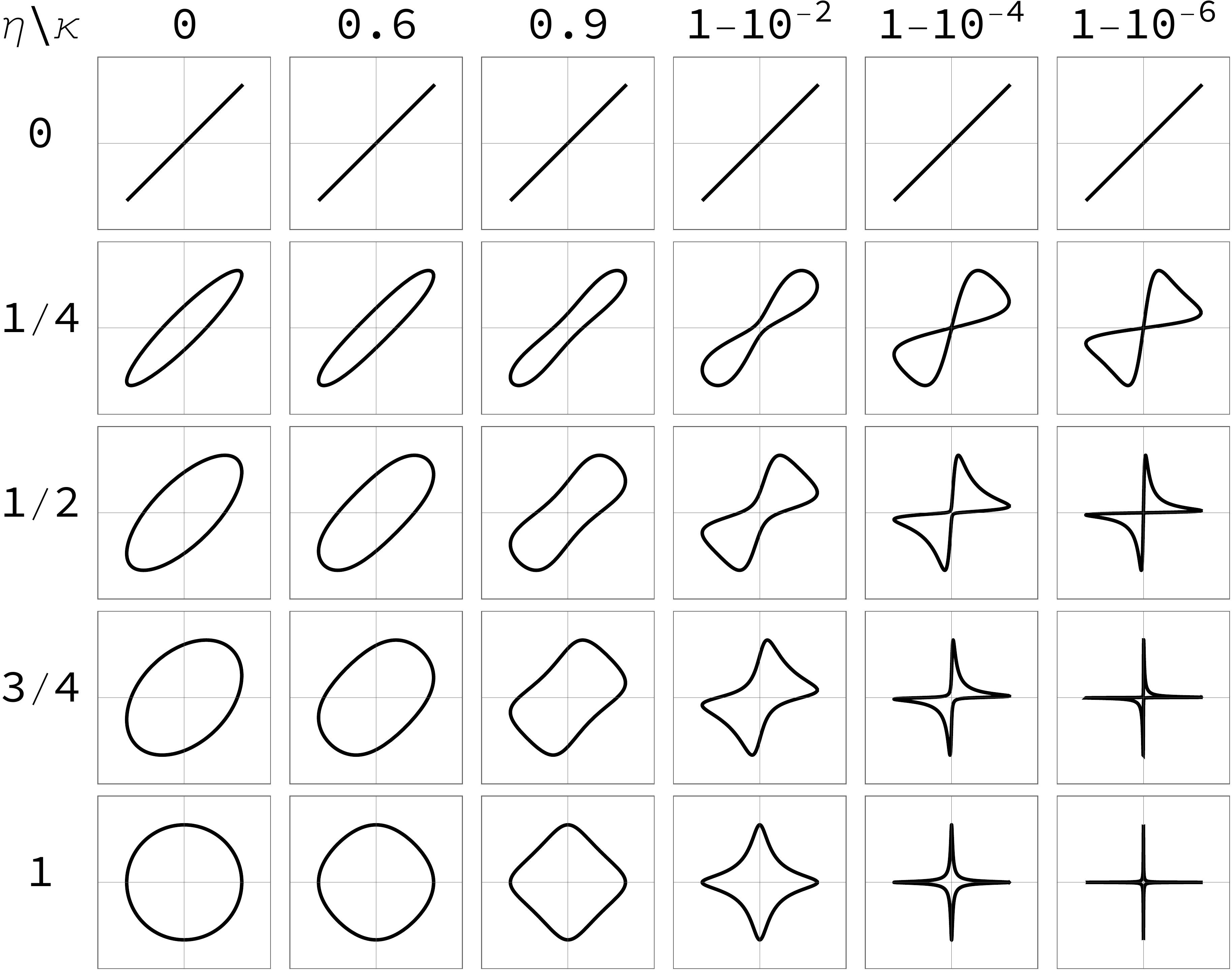}
    \caption{\label{fig:LssjCN}
    Lissajous-Jacobi figures ($\cn$ vs. $\cn$) for a unit ratio,
    $\omega_1:\omega_2 = 1$.
    Columns and rows corresponds to different values of elliptic
    modulus $\kappa$ and phase difference $\eta$.
    Note that $\eta$ is measured in units of quarter period
    $\mathrm{K}[\kappa]$.
    }
\end{figure}

\begin{figure}[t!]
    \centering
    \includegraphics[width=0.7\linewidth]{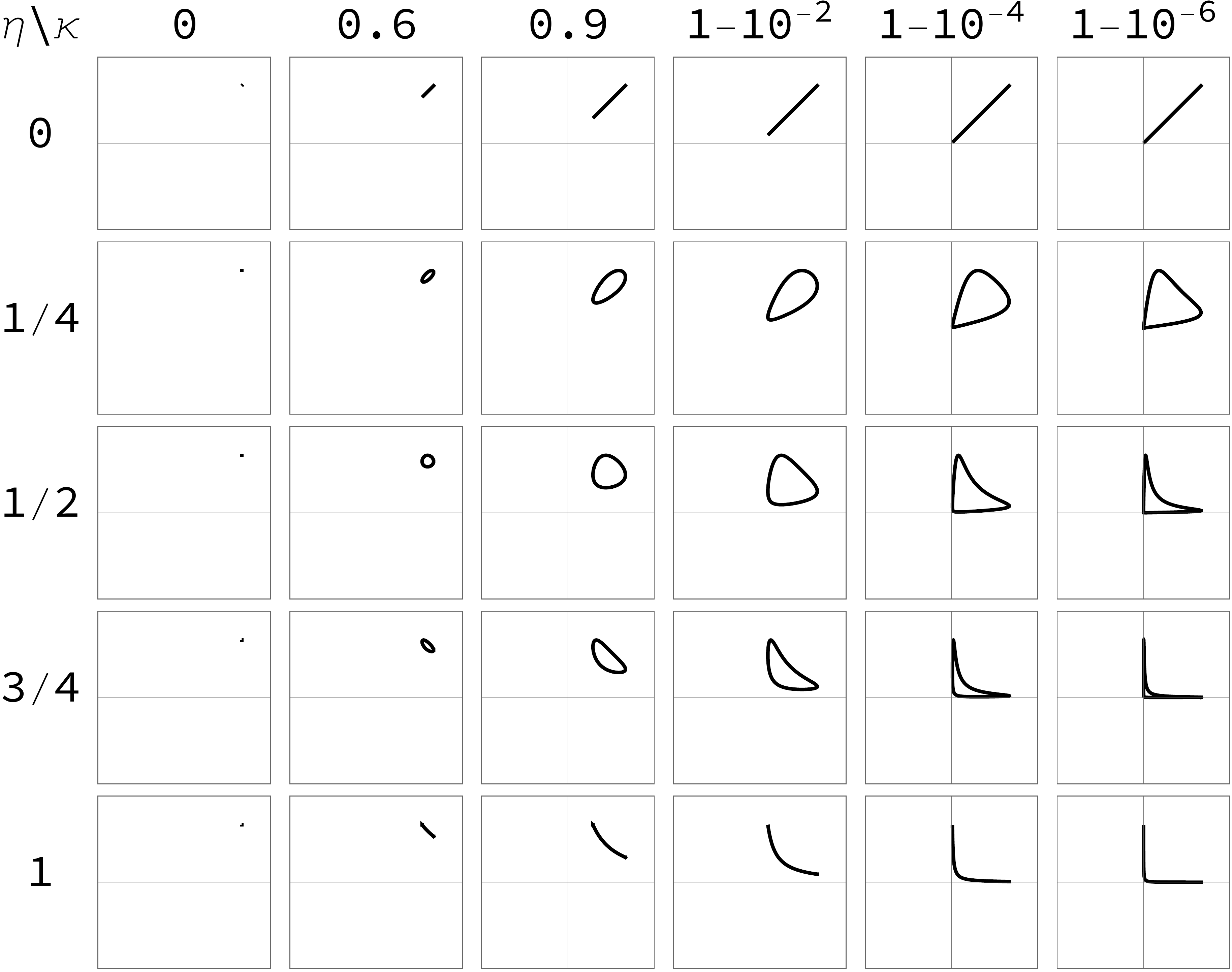}
    \caption{\label{fig:LssjDN}
    Lissajous-Jacobi figures ($\dn$ vs. $\dn$) for a unit ratio,
    $\omega_1:\omega_2 = 1$.
    Columns and rows corresponds to different values of elliptic
    modulus $\kappa$ and phase difference $\eta$.
    Note that $\eta$ is measured in units of quarter period
    $\mathrm{K}[\kappa]$.
    }
\end{figure}

In this article we are mainly interested in closed curves with unit
ratio, $\omega_1:\omega_2=1$, and especially the area they enclose.
Area can be found using the integral $S=\oint y\,\dd x$ resulting in
\begin{eqnarray*}
S_{\text{sn}} & = &
\frac{4}{\kappa^2}\frac{1}{\sn^3 \eta}\Big\{
	- \dn^2 \eta                \,\mathrm{K}[\kappa]
	+ \sn^2 \eta                \,\mathrm{E}[\kappa]
	+ \cn^2 \eta\,\dn^2 \eta    \,\Pi[k^2\sn^2 \eta,\kappa]
\Big\},
\\[0.3cm]
S_{\text{cn}} & = &
\frac{4}{\kappa^2}\frac{\dn\,\eta}{\sn^3 \eta}\Big\{
	                              \mathrm{K}[\kappa]
	- \sn^2 \eta                \,\mathrm{E}[\kappa]
	- \cn^2 \eta                \,\Pi[k^2\sn^2 \eta,\kappa]
\Big\},
\\[0.3cm]
S_{\text{dn}} & = &
\,\,2\,\,\frac{\cn\,\eta}{\sn^3 \eta}\Big\{
	  (\sn^2 \eta+\dn^2 \eta)   \,\mathrm{K}[\kappa]
	- \sn^2 \eta                \,\mathrm{E}[\kappa]
	- \dn^2 \eta                \,\Pi[k^2\sn^2 \eta,\kappa]
\Big\}.
\end{eqnarray*}
Figure~\ref{fig:LssjArea} illustrates behavior of area as a function
of $\eta$ and $\kappa$.
Figures~\ref{fig:LssjSN} -- \ref{fig:LssjDN} show examples of
Lissajous-Jacobi figures for major elliptic functions $\sn$, $\cn$
and $\dn$.
For all plots $\eta$ is measured in $\mathrm{K}[\kappa]$ known as the
quarter period and being the complete elliptic integral of the first
kind.

\newpage

\newpage


%

\end{document}